\documentclass[prd,twocolumn,showpacs,preprintnumbers,nofootinbib,amsmath,amssymb]{revtex4}

\usepackage{graphicx}
\usepackage{dcolumn}
\usepackage{bm}
\usepackage{amsmath}
\usepackage{longtable}

\newcommand{\beq}{\begin{eqnarray}}
\newcommand{\eeq}{\end{eqnarray}}

\newcommand{\real}{{\sf I}\kern-.12em{\sf R}}
\newcommand{\comp}{{\sf I}\kern-.50em{\sf C}}
\newcommand{\unity}{{\sf I}\kern-.54em{\sf 1}}

\def\spose#1{\hbox to 0pt{#1\hss}}
\def\ltapprox{\mathrel{\spose{\lower 3pt\hbox{$\mathchar"218$}}
 \raise 2.0pt\hbox{$\mathchar"13C$}}}

\begin{document}


\title{Thermodynamics of two flavor QCD from imaginary chemical potentials}

\author{Massimo D'Elia $^1$}
\author{Francesco Sanfilippo $^2$}
\affiliation{$^1$Dip. di Fisica, Universit\`a
di Genova and INFN, Via Dodecaneso 33, 16146 Genova, Italy\\
$^2$Dip. di Fisica, 
Universit\`a di Roma ``La Sapienza'' and INFN, P.le A. Moro 5, 00185 Roma, Italy}

\date{\today}

\begin{abstract}
We study QCD thermodynamics in presence of two independent imaginary chemical
potentials coupled to two degenerate flavors of staggered quarks. Analytic
continuation is used to determine non-linear susceptibilities,
to test the Hadron Resonance Gas (HRG) model below
the zero density critical temperature, $T_c$, 
and to determine the average phase factor of the 
fermion determinant. Deviations from HRG predictions, of the 
order of a few percent, are clearly visible for temperatures 
$T > 0.95\ T_c$. The determination of non-linear susceptibilities, 
using different interpolating functions for analytic continuation,
gives consistent results and in agreement with Taylor expansion computations,
apart from some systematic effects at or right above $T_c$.
Results for the average phase factor are compared with the predictions
of Chiral Perturbation Theory; below $T_c$ we are able to
distinguish the contribution of different hadron states, which is 
positive (i.e. tends to mitigate the sign problem) in the case of baryons. 
\end{abstract}


\maketitle

\section{Introduction}
\label{intro}

The study of QCD at finite temperature and baryon density 
has increasing phenomenological interest related to the physics
of heavy ion experiments and compact astrophysical objects.
The main open questions regard the location and nature
of phase transitions in the QCD phase diagram, 
as well as the properties of strongly interacting matter around 
the transitions. A reliable answer to these questions requires
at treatment of QCD at a non-perturbative level: unfortunately 
lattice QCD simulations, which are the only available tool 
for a non-perturbative study of the theory
based on first principles, are not possible at finite baryon chemical 
potential, because of the well known {\em sign problem}:
the QCD fermion determinant becomes complex and the 
probability interpretation of the QCD Euclidean action, 
necessary for standard importance sampling Monte-Carlo, is lost.

A number of strategies have been developed to partially circumvent
that problem, like 
reweighting techniques~\cite{glasgow,fodor,density},
the use of an imaginary chemical potential either 
for analytic continuation~\cite{muim,immu_dl,azcoiti,chen,giudice,cea,sqgp,conradi,cea2} 
or for reconstructing the canonical
partition function~\cite{rw,cano1,cano2}, Taylor expansion
techniques~\cite{taylor1,taylor2,gagu1,gagu2,gagu3} and
non-relativistic expansions~\cite{hmass1,hmass2,hmass3}.

The aim of the present work is that of exploiting the method of analytic
continuation from an imaginary chemical potential to study the
properties of hadronic matter around the deconfinement transition 
in QCD with two light flavors ($N_f = 2$). As an improvement with
respect to previous studies based on analytic continuation, we
introduce to independent chemical potentials, $\mu_1$ and $\mu_2$, 
coupled to the two different quark flavors. 
That is equivalent to the introduction of two independent chemical 
potentials, $\mu_B$ and $\mu_I$, coupled respectively to the baryon
and to the isospin charges $B$ and $I_3$.

Our strategy will be to determine the dependence of the free energy on
the two chemical potentials, apart from constant terms, 
by measuring its first derivatives with respect to $\mu_1$ and $\mu_2$
(quark number densities) for imaginary values of the two variables, 
and by then fitting them by suitable functions, to be continued 
within proper analyticity domains.

One of our aims is the study of generalized
susceptibilities with respect to different conserved charges of the
model (baryonic, isospin). These quantities are of significant
phenomenological interest
and have been
determined till now mostly by the Taylor expansion method. We shall
compare our results with those obtained by previous studies and 
comment on the efficiency and systematic effects of analytic continuation.
In the confined region, i.e. below the critical temperature $T_c$, we shall
be able to perform a high precision test of the Hadron Resonance Gas 
(HRG) model, leading to the uncover of violations close to $T_c$.
Finally, the knowledge of the dependence of the free energy on the 
two independent chemical potentials will allow us a study of the 
average phase factor, which gives a direct measurement of the
severeness of the sign problem.

Our study is made for QCD with two flavors of unimproved staggered quarks
and is based on a standard RHMC algorithm.
The choice of parameters is taken from Ref.~\cite{gagu2}.
The paper is organized as follows:
In Section II we describe the model that we have
investigated as well as the relevant physical observables; we also discuss the
symmetries of the model, which are important for
the choice of the free energy interpolating functions to be used for analytic
continuation.
In Section III we report the technical details of our numerical
simulations. In Section IV we present results obtained below 
$T_c$ and compare them to the predictions of the HRG model.
In Section V we report results obtained above $T_c$. 
In Section VI and VII we discuss results obtained respectively
for generalized susceptibilities and for the analytic continuation
of the average phase factor.
Finally, in Section VII, we draw our conclusions.

\section{$N_f = 2$ QCD with two independent chemical potentials and 
analytic continuation.}
\label{gensusc}

QCD with two continuum degenerate flavors is described, in the
(rooted) staggered fermion discretization of the theory, by the following
partition function
\beq
Z(T) \equiv \int \mathcal{D}U e^{-S_{G}[U]} (\det M[U])^{1/2} 
\eeq
where $S_G$ is the discretized pure gauge action (standard 
Wilson plaquette action in our case) and $M$ is the staggered fermion 
matrix describing 4 continuum flavors. Periodic (antiperiodic) 
boundary conditions are assumed for gauge (fermion) fields along
the Euclidean time direction.

The introduction of two independent chemical potentials, $\mu_1$ and 
$\mu_2$, coupled to the number operators of each quark family leads
to the following expression for the grand canonical partition function:
\beq
Z(T,\mu_1,\mu_2) \equiv \int \mathcal{D}U e^{-S_{G}} 
\det M^{1\over 4} [\mu_1]
\det M^{1\over 4} [\mu_2]
\label{partfun1}
\eeq
where the fermion matrix in the standard staggered formulation at finite
chemical potential reads:
\begin{eqnarray}
M[\mu]_{i,j} &=& a m
\delta_{i,j} + {1 \over 2} 
\sum_{\nu=1}^{3}\eta_{i,\nu}\left(U_{i,\nu}\delta_{i,j-\hat\nu}-
U^{\dag}_{i-\hat\nu,\nu}\delta_{i,j+\hat\nu}\right) \nonumber \\
&+& \eta_{i,4}
\left(e^{ a \mu}U_{i,4}\delta_{i,j-\hat4}-
e^{- a \mu}U^{\dag}_{i-\hat4,4}\delta_{i,j+\hat4}\right) 
\label{fmatrix}
\end{eqnarray}
Here $i$ and $j$ refer to lattice sites, $\hat\nu$ is a unit vector on
the lattice, $\eta_{i,\nu}$ are staggered phases;
$a \mu$ and $a m$ are respectively 
the chemical potential and the quark mass in lattice units.

The two chemical potentials can be rewritten
in terms of a quark number chemical potential 
$\mu_q = (\mu_1 + \mu_2)/2$ (or equivalently a baryon chemical
potential $\mu_B = 3 \mu_q$) and of an isospin chemical potential  
$\mu_I = (\mu_1 - \mu_2)/2$.

While the original theory is invariant under both charge conjugation
and isospin rotations, the theory in presence of finite chemical potentials
obviously is not. However the original invariance
is reflected in the fact that the free energy 
$F = - T \ln Z $
must be an even function of $\mu_q$ and $\mu_I$ separately, or equivalently 
it must be invariant under the two following transformations
$(\mu_1,\mu_2) \to (\mu_2,\mu_1)$ and 
$(\mu_1,\mu_2) \to (-\mu_2,-\mu_1)$, which are easily verified to be 
symmetries of the partition function in Eq.~(\ref{partfun1}). 
That places strong constraints on its possible functional dependence.

In presence of a finite chemical potential $\det M$ becomes
complex and $\det M[-\mu] = (\det M[\mu])^*$. Therefore,
apart from the case $\mu_2 = - \mu_1$ ($\mu_q = 0$), the integrand
in Eq.~(\ref{partfun1}) is complex and cannot be interpreted as 
a probability distribution over gauge fields, so that standard 
importance sampling techniques cannot be applied (sign problem).

Positivity is recovered if the chemical potentials $\mu_1$ and 
$\mu_2$ are taken as purely imaginary: in this case numerical simulations
are feasible and results can be used to fit the functional dependence
of relevant observables.

Due to the above mentioned symmetries of the free energy,
analytic continuation is actually a continuation from negative to
positive values of $\mu_q^2$ and $\mu_I^2$. Of course 
it is expected to be applicable as long as no phase transitions 
are met along the continuation path.

It is convenient for the following discussion to introduce 
the variables 
$$\theta_q = {\rm Im}(\mu_q)/T = N_t a {\rm Im}(\mu_q) $$
$$\theta_I = {\rm Im}(\mu_I)/T = N_t a {\rm Im}(\mu_I) $$ 
and 
$$\theta_1 = {\rm Im}(\mu_1)/T = \theta_q + \theta_I$$
$$\theta_2 = {\rm Im}(\mu_2)/T = \theta_q - \theta_I$$
where $N_t$ is the number of lattice sites in the temporal direction.

It can be easily shown that the introduction of 
an imaginary chemical potential is equivalent to a twist 
in the temporal boundary conditions for fermions by an angle 
${\rm Im} (\mu)/T$. Hence
both determinants appearing in Eq.~(\ref{partfun1}) are
periodic functions, respectively of $\theta_1$ and $\theta_2$, with
period $2 \pi$, so that the free energy itself is a periodic function 
of these variables.

In terms of $\theta_q$ and
$\theta_I$ that means again periodicity with period $2 \pi$ in both
variables, plus invariance under 
$(\theta_q,\theta_I) \to (\theta_q + \pi,\theta_I + \pi)$.
However, following the argument given by Roberge and Weiss 
in Ref.~\cite{rw}, it is possible to prove that a transformation 
$\theta_q \to \theta_q + 2 \pi k/N_c$, where $N_c$ is the number of colors
and $k$ is an integer, 
can be cancelled by a change of variables in the functional
integration in which all temporal links at a given time slice get
multiplied by a center element $\exp(-i 2 k \pi/N_c)$ (center
transformation). 
Hence the free energy is expected to be a periodic function of $\theta_q$ with 
period $2 \pi/N_c$ instead of $2 \pi$ ($N_c = 3$ in our case). 
An analogous change of variables 
does not work for translations in $\theta_I$, which rotate the link
variables appearing in each determinant in a different way, therefore 
the period in $\theta_I$ is really $2 \pi$.

For temperatures below the zero density critical temperature, $T_c$,
no phase transitions are expected, as in the $\mu_I = 0$ case, 
in the whole $\theta_q,\theta_I$ plane. Therefore, due to the 
discussed periodicity and required symmetries, 
the most natural parametrization of the free energy is in terms of a 
trigonometric series as follows:
\beq
\frac{F(\theta_q,\theta_I)}{T} = \sum_{h,l} w_{h,l} \cos (3 h \theta_q) \cos(l \theta_I)
\label{freelowT}
\eeq
with $h$ and $l$ both integers; moreover $h$ and $l$ must have the same parity
because of the invariance under 
$(\theta_q,\theta_I) \to (\theta_q + \pi,\theta_I + \pi)$.
Further constraints on the number of terms appearing in Eq.~(\ref{freelowT})
may be predicted by particular effective models of strong interactions
below $T_c$, like for instance the HRG model to be discussed in
Section~\ref{resultsHRG}. In such regime, information valid for
analytic continuation can be gathered in the whole $\theta_q,\theta_I$ plane.

For $T > T_c$ we expect instead phase transitions in the 
$\theta_q,\theta_I$ plane, corresponding either to the continuation
of the physical deconfinement transition or to the generalization of 
Roberge-Weiss (RW) transitions.
Therefore a limited region around
$\theta_q = \theta_I = 0$ is available for the purpose of analytic
continuation to real chemical potentials, and 
we shall write an expression for the free energy valid in that region
which respects the predicted symmetries under 
$\theta_q \to - \theta_q$ and $\theta_I \to -
\theta_I$ separately. In particular the free energy will be expressed 
as a polynomial like
\beq
\frac{F(\theta_q,\theta_I)}{T} = \sum_{i,j} c_{i,j} 
\frac{\theta_q^{2 i}}{(2i)!}\frac{\theta_I^{2 j}}{(2j)!}
\label{freehighT1}
\eeq
with $i,j$ non negative integers,
or as a ratio of polynomials of the same kind 
\beq
\frac{F(\theta_q,\theta_I)}{T} = 
\frac
    {\left.\sum_{i,j} n_{i,j} 
      \frac{\theta_q^{2 i}}{(2i)!} \frac{\theta_I^{2 j}}{(2j)!}
      \right|_{n_{00}=0}}
    {\left.\sum_{k,l} d_{k,l} 
      \frac{\theta_q^{2 k}}{(2k)!} \frac{\theta_I^{2 l}}{(2l)!}
    \right|_{d_{00}=1}} \,.
\label{freehighT2}
\eeq
The latter is an example of Chisholm approximant, i.e. the
generalization to the case of two independent variables 
of usual Pad\`e approximants, 
which have revealed to be better suited for analytic 
continuation in some cases~\cite{mpl05,cea,shinno}.

Some of the quantities we are interested in are generalized 
susceptibilities with respect to the different chemical
potentials, which for $N_f = 2$ are defined as follows
\beq
\chi_{i,j} \equiv \frac{\partial^{i+j}}{\partial \mu_1^i \partial \mu_2^j}
\left( -\frac{F}{V} \right) = 
\frac{\partial^{i+j}}{\partial \mu_1^i \partial \mu_2^j}
P \label{gensusc1}
\eeq 
where $P$ is the pressure. Analogous susceptibilities are
defined in terms of $\mu_q$ and $\mu_I$
\beq
\chi_{i,j}^{q,I} \equiv \frac{\partial^{i+j}}{\partial \mu_q^i \partial \mu_I^j}
P \label{gensusc2} \, .
\eeq 
The free energy symmetries discussed above imply precise
constraints on the susceptibilities computed at zero 
chemical potentials. In particular we have
$\chi_{i,j}^{q,I} \neq 0$ only if $i$ and $j$ are both even,
while $\chi_{i,j} \neq 0$ if $i + j$ is even and 
$\chi_{i,j} = \chi_{j,i}$.

Such quantities encode all relevant information about fluctuations of 
conserved charges, which are generally considered to be sensitive
probes for the properties of the thermal medium produced in heavy
ion collisions. 
They have been computed mostly in the Taylor expansion 
approach~\cite{taylor1,taylor2,gagu1,gagu2,gagu3}, 
where they are expressed as average values at $\mu = 0$
of operators which are more and more complex and computationally
demanding as the order grows, since they require more and more 
matrix inversions. It is therefore sensible to 
explore the consistency and the efficiency of different strategies. In the
analytic continuation approach we  
determine numerically the functional dependence,
for imaginary values of the chemical potentials, 
of the first derivatives of the free energy. In terms of 
adimensional quantities, which are most conveniently determined on the
lattice, they are given by
\beq
\hat{n}_q &\equiv& \frac{\langle N_q \rangle}{V T^3}  = 
 \frac{\partial}{\partial \mu_q} (P/T^3) = \hat{n}_1 + \hat{n}_2
\nonumber \\
\hat{n}_{I} &\equiv&  \frac{\langle N_{I} \rangle}{V T^3}  = 
 \frac{\partial}{\partial \mu_I} (P/T^3) = 
\hat{n}_1 - \hat{n}_2
\label{numbers1}
\eeq
where $N_q$ and $N_{I}$ are respectively the quark number and
isospin charge operators, with
\beq
\hat{n}_i &\equiv& \frac{\langle N_i \rangle}{VT^3} = \frac{1}{VT^2} 
\frac{\partial \ln Z}{\partial \mu_i} = -
\frac{1}{VT^3} \frac{\partial F }{\partial \mu_i}  \nonumber \\
&=& \frac{N_t^2}{4 N_s^3} 
\left\langle {\rm Tr} \left( M^{-1}[U,\mu_i] \frac{\partial}{\partial
  a \mu_i} M[U,\mu_i] \right) \right\rangle 
\label{numbers2}
\eeq
for $i = 1,2$.
In terms of the susceptibilities defined in Eq.~(\ref{gensusc1}) 
$\hat{n}_1 = \chi_{1,0}/T^3 $ and $\hat{n}_2 = \chi_{0,1}/T^3 $.
Such first derivatives, which are purely imaginary for imaginary
chemical potentials, can be measured quite efficiently
(only one matrix inversion is needed for the noisy estimation of the
trace) and, apart from constant terms, 
encode all information about the dependence of the free energy
on $\mu_q,\mu_I$. 
Information gathered at imaginary values of $\mu_{q/I}$ can then be 
analytically continued to real values of $\mu_{q/I}$, in particular 
higher order derivatives at $\mu_q = \mu_I = 0$ can be extracted.

In comparison to the Taylor expansion approach, the 
great advantage related to the much simpler observables can be 
compensated by the need for multiple simulations at different
values of the chemical potentials. Moreover, 
this procedure involves some systematic dependence on the 
function chosen to interpolate data at imaginary $\mu$'s, which should
be eventually checked by comparing results obtained with different functions.
We shall compare trigometric expansions with polynomials below $T_c$, 
polynomials with ratio of polynomials above $T_c$.

\section{Parameter details and numerical setup}
\label{setup}

Since we want to compare our results for the 
generalized susceptibilities with those obtained by the Taylor
expansion approach, we have chosen for this study a subset of
the parameters used in Ref.~\cite{gagu2}, which is reported
in Table~\ref{tabPARsim}. That corresponds to 
five different temperatures with a standard staggered lattice
discretization on $N_t = 4$ lattices and a fixed value (on 
the corresponding $T = 0$ lattices) for the pion mass, 
$m_\pi \simeq 280$ MeV (actually $m_\pi/m_\rho = 0.31(1)$ and
$m_\rho/T_c = 0.54(2)$).  The critical temperature reported
in Ref.~\cite{gagu2} is $T_c \simeq 170$ MeV.

\begin{table}
\begin{center}
\begin{tabular}{|c||c|c||c|c|c|}
\hline $T/T_c$ & $m_q$ & $\beta$ & $n_{pairs}$ & $n_{traj}$ & $N_{D}$\\
\hline
\hline 0.9   & 0.02778 & 5.26   & 95 & 2300 & $12.6 \cdot 10^{9}$ \\
\hline 0.951 & 0.02631 & 5.275  & 95 & 2460 & $14.0 \cdot 10^{9}$ \\
\hline 1     & 0.025   & 5.2875 & 95 & 3500 & $20.7 \cdot 10^{9}$ \\
\hline 1.048 & 0.0238  & 5.30   & 24 & 3120 & $4.7  \cdot 10^{9}$ \\
\hline 1.25  & 0.02    & 5.35   & 77 & 2270 & $8.7  \cdot 10^{9}$ \\
\hline
\end{tabular}
\end{center}
\caption{List of simulated temperatures and corresponding $\beta,m_q$ values 
(taken from Ref.~\cite{gagu2}), number of $(\mu_q,\mu_I)$ pairs explored at 
each temperaure ($n_{pairs}$) and average number of trajectories (of 1
MD time length each) generated 
at each temperature and for each $(\mu_q,\mu_I)$ pair ($n_{traj}$).
$N_{D}$ instead indicates the total number of Dirac matrix
multiplications performed at each $T$, 
which is reported as an estimate of the overall computing effort
performed:
that is more or less equally distributed between Monte-Carlo and 
measurements.
\label{tabPARsim}}
\end{table}

In particular, we have made simulations on a $16^3 \times 4$ lattice using a 
RHMC algorithm. Our spatial size $L_s = 16$ corresponds to about 6.6
inverse pion masses, hence finite size effects are not
expected to be important.

For $T \leq T_c$ we have made simulations on a grid of 
about 100 different pairs
$(\theta_q,\theta_I)$, in the range $[0,\pi] \times [0,\pi]$: because of the 
above described periodicity, this surely contains all possible information 
available at imaginary chemical potentials (actually in a redundant
way, which 
however is a benefit for checking the reliability of our statistical analysis).
Since susceptibilities are calculated at null values of $\mu_q$ and 
$\mu_I$, more points were taken in a 
restricted region around the origin, in order to perform fits of 
low-degrees polynomials in $n_q$ and $n_I$ around 
the origin easily.
Morover, we have decided to perform a more accurate study of HRG model along
the axis $\theta_I = 0$, therefore we have chosen further points there.

For $T > T_c$ we have performed a preliminary study aimed at finding the position of 
transition lines, with the purpose of delimiting the region at imaginary
chemical potentials available for analytic continuation.
Further information about this region are given in 
Section~\ref{formaREGIONEbuona}.

For each $(T,\theta_q,\theta_I)$ we have produced about 2-3K thermalized 
trajectories of 1 Molecular Dynamics time length each. 
More details about the amount of $(\mu_q,\mu_I)$ pairs explored
and average numbers of generated configurations are 
given in Table~\ref{tabPARsim}.

Quark densities have been measured by using noisy estimators. It is possible to
minimize the total error of these observables (sum of statistical and noise
fluctuations) at fixed simulation time by choosing an appropriate number of
random vectors used for each noisy estimation. Assuming that noise and
statistical fluctuations are independent of each other, the optimal number of 
random vectors $n_{vec}$ to be used for each configurations is given by 
\beq
n_{vec}=\frac{\sigma_{noise}}{\sigma_{meas}}
\sqrt{\frac{\tau_{conf}}{\tau_{estim}}}
\label{optimalnvec}
\eeq
where $\sigma_{meas}$ is the variance of the observable (quark density) over 
different configurations, $\sigma_{noise}$ is the variance of the different 
estimates of the observable over a fixed configuration, $\tau_{conf}$ is the 
time needed to generate a new configuration and $\tau_{estim}$ is the time 
needed to perform one noisy estimate of the observable. We have
measured those quantities in preliminary runs and we have found that,
with our numerical setup, this number is around 30 for all
explored parameter sets. Notice that Eq.~(\ref{optimalnvec})
does not take into account the autocorrelation among configurations
and thus overestimates $n_{vec}$; we have however directly checked, 
by comparing different choices of $n_{vec}$, that the efficiency is 
almost stable for $n_{vec} \sim 10-50$. We have always chosen 
$n_{vec} = 30$ in our production runs.

Simulations have been done on two PC farms in Genoa and in Bari.
The complete collection of our data is not reported here, but is 
at disposal for interested readers.

\section{Results at $T \leq T_c$: precision test on the Hadron
  Resonance Gas model}
\label{resultsHRG}

The thermal medium below the critical temperature is generally believed
to be well described as a gas of free hadron resonances (HRG model).
This model provides a good description of thermal conditions at 
freeze-out~\cite{redlich,redlich2,andronic} and has 
received theoretical support from lattice QCD 
simulations~\cite{kareta}. Deviations from
the model have been recently detected close to $T_c$ 
in a lattice study based on the Taylor expansion method~\cite{taylor2}.

In the HRG model the free energy is expressed as the sum of free
particle energies. In particular, the free energy for species $i$ 
of spin $s_i$, mass $m_i$,
baryon number $B_i$ and isospin $I_{3i}$, is given by 
\beq
- T \ln Z_i &=& \pm \frac{g_i V T}{2 \pi^2} \int_0^\infty
\ln \left( 1 \mp z_i e^{\frac{\sqrt{m_i^2 + k^2}}{T}} \right) k^2 dk
\nonumber \\
&=& \frac{g_i VT^2 m_i^2}{2 \pi^2} \sum_{l = 1}^{\infty}
\left[ \frac{(\pm 1)^{l + 1}}{l^2}z_i^l K_2 \left(\frac{m_i l}{T}
  \right) \right]
\label{freeparticle}
\eeq
where $g_i = 2 s_i + 1$, the upper (lower) sign applies to mesons
(baryons) and 
\beq
z_i = e^{\mu_i/T} = \exp\left( \frac{3 B_i \mu_q + 2 I_{3i}
  \mu_I}{T} \right) \, .
\eeq
The expression in Eq.~(\ref{freeparticle}) is an approximation
in the case of unstable particles, for which an integration over a Breit-Wigner
distribution in the particle mass would be more appropriate. 
The Bessel function $K_2$ is exponentially suppressed for large
values of the argument, $K_2(x) \simeq \sqrt{\pi / (2 x)} e^{-x}$,
hence for $m_i \gg T$ we can keep just the first
term $l = 1$ in the $l$ expansion, corresponding to the Boltzmann
approximation in which quantum statistics effects are neglected. 
Summing up over all known particles and resonances 
and grouping together all charge conjugation and isospin 
partners we get
\beq
\ln Z = V T^3 && \sum_{B,I,m}  W(m,g,T) 
\bar\delta(B)
\cosh \left( 3 B
\frac{\mu_q}{T} \right) \nonumber \\
&& \left( \sum_{I_3 \geq 0} 
\bar\delta(I_3)\cosh \left( 2 I_3
\frac{\mu_I}{T} \right) \right)
\eeq
where $\bar\delta (n) = 1 - {1}/{2} \delta_{n,0}\,\,$ and 
$$W(m,g,T) = 2g \left( m\over {\pi T} \right)^2
K_2 \left( m\over T \right) \, .$$
Such prediction is easily continued to imaginary chemical potentials,
where hyperbolic functions get transformed into trigonometric functions,
in particular we have
\beq
\ln Z = V T^3 &&\sum_{B,I}  W_{B,I}(T) 
\bar\delta(B)
\cos ( 3 B \theta_q) \nonumber \\ && \left( \sum_{I_3 \geq 0} 
\bar\delta(I_3)\cos ( 2 I_3 \theta_I ) \right)
\label{imF}
\eeq
\beq
 {\rm Im} (\hat{n}_{q}) 
= 
&&\sum_{B,I}  3 B W_{B,I}(T) 
\sin ( 3 B \theta_q) \nonumber \\ && \left( \sum_{I_3 \geq 0} 
\bar\delta(I_3)\cos ( 2 I_3 \theta_I ) \right)
\label{imNq}
\eeq
\beq
{\rm Im}( \hat{n}_{I}) 
=
&&\sum_{B,I} W_{B,I}(T) 
\bar\delta(B)
\cos ( 3 B \theta_q) \nonumber \\ && \left( \sum_{I_3 \geq 0} 
2 I_3 \sin ( 2 I_3 \theta_I ) \right) 
\label{imNi}
\eeq
where $W_{B,I}(T) = \sum_{m|_{B,I}} W(m,g,T)$.
The average quark densities are always purely imaginary for 
imaginary chemical potentials, for that reason we shall simply write
$\hat{n}_{q}$ and $\hat{n}_{I}$ in the following, meaning implicitely
that their imaginary part is taken.

Predictions from the HRG model to be tested in lattice QCD simulations
can be classified as follows:

1) The free energy has a particularly simple form since, on the basis
of known hadron resonances, only
$W_{0,0}$, $W_{0,1}$, $W_{1,1/2}$, $W_{1,3/2}$ are different from
zero in previous equations. That means a further strong restriction 
on the expected form of the free energy at low temperatures:
a necessary condition for the HRG model to be valid
is that only the few lowest terms of the Fourier expansion 
in Eq.~(\ref{freelowT}) give contribution;

2) Also the numerical values of the coefficients can be predicted from
the known experimental resonance mass spectrum.

Latter prediction is easily affected by lattice artifacts
and by the unphysical quark masses used in simulations, which 
change the actual hadron spectrum on the lattice. The former, instead,
is expected to be more robust and less sensitive to discretization
details. The method of analytic continuation is particularly 
well suited for lattice QCD tests of the HRG model, since it gathers
information, below $T_c$, from the whole range of possible imaginary
chemical potentials, so that the number of terms actually contributing  
to the Fourier expansion in Eq.~(\ref{freelowT}) can be checked
with great precision: this idea has been followed in earlier studies
limited to the $\theta_I = 0$ axis~\cite{immu_dl,cano1}, 
in which the presence, within errors, of a single Fourier contribution, 
corresponding to $B = 1$, has been verified. In this respect the aim of 
our work is to extend such studies by increasing precision and by 
exploring also the $\theta_I \neq 0$ region.

\begin{figure}[!ht]
\includegraphics*[width=1.0\columnwidth]{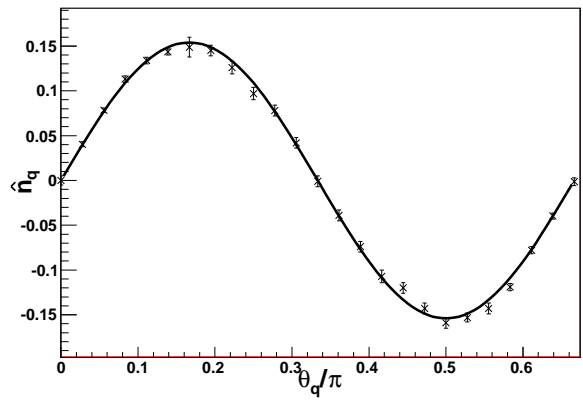}
\vspace{-0.cm}
\caption{Normalized quark density at $T = 0.9\, T_c$ and $\theta_I =
  0$. The solid line corresponds to the single sine fit reported in 
Table~\ref{tabSINfit}.}
\label{figSINfitASSEb0.9} 
\vspace{-0.cm}
\end{figure}

\subsection{$T = 0.9\, T_c$}
\label{T0.9}

We start by discussing results obtained at $T = 0.9\, T_c$. Let us
first look at the $\theta_I = 0$ axis: $\hat{n}_I$ is zero in this
case, while in general $\hat{n}_q$ can be Fourier expanded as:
\beq
\hat{n}_q =\sum_{l=1} c_l \sin\left(3 l \theta_q\right)
\label{sinfit}
\eeq
and the HRG model predicts contribution only from the lowest harmonic,
$l = 1$. 
Indeed a simple sine term,
corresponding to $B = 1$, is perfectly compatible with our data,
as showed in Fig.~\ref{figSINfitASSEb0.9} and reported in
Table~\ref{tabSINfit}. A second term with $l = 2$ is therefore not necessary, 
at least within the precision of our data, even if a two sine fit
leads to a smaller $\chi^2/d.o.f.$ with a $c_2 \neq 0$ within three
standard deviations (see again Table~\ref{tabSINfit}). 
As shown in Table~\ref{tabSINfourier}, completely 
equivalent results are obtained if, instead of fitting our data, 
we compute the coefficients $c_l$ by explicit Fourier transform,
\beq
c_l = \frac{3}{\pi} \int_0^{2 \pi /3} \sin (3 l \theta_q) \hat{n}_q
(\theta_q) d\theta_q
\eeq
where the integration is performed numerically by linear interpolation
of consecutive data points.

\begin{figure}[t!]
\includegraphics*[width=1.0\columnwidth]{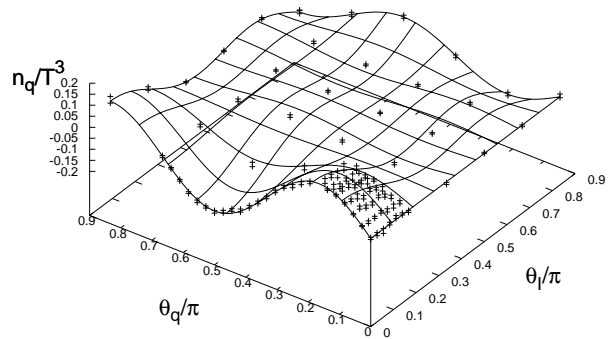}
\vspace{-0.cm}
\caption{Fit of normalized quark densities at $T = 0.9 \, T_c $, obtained 
from all imaginary chemical potentials explored (cross points),
with the prediction from the HRG model (grid surface)}
\label{figHRGfitTOT0.9} 
\vspace{-0.cm}
\end{figure}

\begin{figure}[t!]
\includegraphics*[width=1.0\columnwidth]{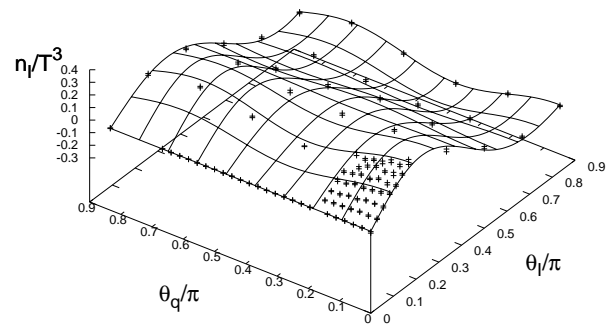}
\vspace{-0.cm}
\caption{Fit of normalized isospin densities at $T = 0.9 \, T_c $, obtained
from all imaginary chemical potentials explored (cross points),
with the prediction from the HRG model (grid surface)}
\label{figHRGfitDIF0.9} 
\vspace{-0.cm}
\end{figure}

Next we consider data for $\hat{n}_q$ and $\hat{n}_I$
obtained in the whole range of $\theta_q$ and $\theta_I$ explored, 
which are shown in Figs.~\ref{figHRGfitTOT0.9} and \ref{figHRGfitDIF0.9}, 
and try to fit them according to the expressions in Eqs.~(\ref{imNq}) and
(\ref{imNi}), considering more and more parameters $W_{B,I}$ till an
acceptable value for the $\tilde{\chi}^2$ test is obtained.
Fit results are reported in Table~\ref{tabHRGfit}: a 
reasonable value of $\tilde{\chi}^2$ is obtained if a term
with quantum numbers $B = 0$ and $I = 2$ is allowed for, besides those
corresponding to usual meson ($B = 0, I = 1$) and baryons 
($B = 1, I = 1/2$ or $3/2$). Such term does not correspond to any known
or even possible exotic hadron~\cite{jaffe}, but it is easily recognized
as the first term, $l = 2$, neglected in Eq.~(\ref{freeparticle}) in
the Boltzmann approximation in the case of pions: this is actually the first
correction taking into account quantum statistics effects for pions,
i.e. the fact that they are bosons, and corresponds to a two-pion exchange.
With a pion mass as that used in our simulations, 
$m_\pi \sim 280$ MeV, such term would mimic a coefficient 
$W_{0,2} \sim 0.0045$, in very good agreement with the value obtained
in our fit. Notice that terms with $l >2$ are negligible 
in our discretization setup, but would not be so, already at this
temperature, in the case of physical pion masses.
As for the data at $\theta_I = 0$, 
allowing for a term with $B = 2$ leads to a lower value of $\tilde{\chi}^2$,
but is not strictly necessary, at least within the precision of our
data.

Our conclusion is therefore that at $T = 0.9 \, T_c$ numerical data
do not contradict, within errors, the prediction coming from the
HRG model and regarding the number of terms actually 
contributing to the free energy, apart from marginal evidence
for a $B = 2$ term which however is not strictly needed to fit data. 
Other deviations can be ascribed to 
the crudeness of the Boltzmann approximation for pions and are indeed 
well accounted for by the first neglected term.

Of course if one looks at the numerical value of the coefficients,
checking the agreement with experimental data is less trivial: 
taking into account all 
non-strange (since we are considering $N_f = 2$) hadron resonances reported 
in the Particle Data Book~\cite{PDG},
we would expect, for instance, 
$W_{0,1} = 0.457$\footnote{More precisely we considered
all mesons of widely accepted existence, marked with a dot in 
the meson summary table.}, 
which is roughly twice the value we have obtained ($W_{0,1} = 0.216(2)$). 
A more careful comparison is made using the unphysical pion and $\rho$ masses 
realized in our lattice simulations ($m_\pi \sim 280$ MeV and 
$m_\rho \sim 918$ MeV~\cite{gagu2}): the coefficient becomes $W_{0,1} \sim 0.30(2)$ 
including all resonances, $W_{0,1} \sim 0.26(2)$ taking into account just
pions and $\rho$ particles, and $W_{0,1} \sim 0.225(15)$ including
just pions (the errors here take roughly into account the
uncertainties given for the lattice estimate of the masses in Ref.~\cite{gagu2}),  
i.e. much closer to our numerical result or even perfectly compatible
in the last case. We notice that, 
since already $\rho$ masses are beyond the UV scale of our lattice
($a^{-1} \sim 700$ MeV), it is perfectly reasonable that the
contribution from higher
resonances is not properly take into account.
That also clearly shows that a comparison of the numerical values
of the fitted coefficients with the HRG model prediction is
unavoidably affected by the systematics of the lattice discretization.

\subsection{$T = 0.951\, T_c$}

\begin{figure}[!ht]
\includegraphics*[width=1.0\columnwidth]{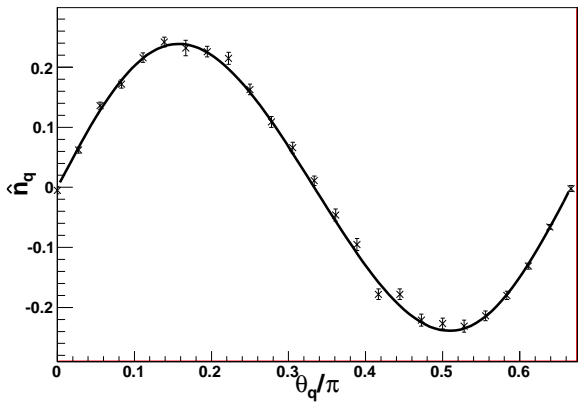}
\vspace{-0.cm}
\caption{Normalized quark density at $T = 0.951\, T_c$ and $\theta_I=0$.
The solid line corresponds to the two-sine fit reported in 
Table~\ref{tabSINfit}.}
\label{figSINfitASSEb0.951}
\vspace{-0.cm}
\end{figure}

Once again we first look at results obtained for $\hat{n}_q$ at $\theta_I = 0$,
which are shown in Fig.~\ref{figSINfitASSEb0.951}.
In this case two Fourier terms, corresponding to 
$B= 1$ and $B = 2$, are necessary to fit our data. The second
term is small, giving a contribution of the order of $5\%$ the 
total signal, but our data are precise enough to detect it;
indeed a $\tilde\chi^2$ of order 2 is obtained if a single sine
fit is tried (see Table~\ref{tabSINfit}).

In this case the presence of the $B = 2$ term cannot be simply ascribed to 
a violation of the Boltzmann approximation: assuming a mass
of order 1 GeV for the lightest baryon, the first neglected
term should lead to a signal a factor
$10^2$ smaller than what we get; moreover it should be negative, 
as appropriate for a two-fermion exchange term. 
The presence in the thermal medium of baryon-baryon bound states,
like deuterons, is a viable hypothesis: however assuming a mass
difference $\Delta M \sim 1$ GeV between those states and 
the lowest baryon states, one would expect a suppression
factor of the order 
$\exp (- \Delta M/T) \sim 10^{-3}$ at this temperature, i.e. much 
smaller than what we have obtained~\footnote{Notice however that also 
for this states lattice artifacts due to the low UV cutoff,
$a^{-1} \sim 700$ MeV, could be important.}.
A simpler explanation is 
that at this temperature corrections to the HRG model, induced by 
non-trivial interactions close to the phase transition,
start to be important.

That is confirmed by analyzing the complete set of data for
$\hat{n}_q$ and $\hat{n}_I$ as a function of
$\theta_q$ and $\theta_I$: fit results are reported in
Table~\ref{tabHRGfit}.
Also in this case a term with $(B,I) = (0,2)$ is needed, but 
its value comes out to be about twice than 
expected from the first term neglected
in the Boltzmann approximation for pions.
In order to get a reasonable value for $\tilde \chi^2$ it is 
necessary to introduce also terms corresponding to $B = 2$
(in agreement with results at $\theta_I = 0$) and terms with 
$B = 1$ and isospin up to $I = 7/2$. We interpret
this again as a violation of the HRG model.

Regarding the numerical values obtained for the fitted coefficients,
we obtain for instance $W_{0,1} \sim 0.256(2)$, to be compared with 
$W_{0,1} \sim 0.24$ if only pions are taken into account,  
$W_{0,1} \sim 0.28$ including pions and $\rho$ mesons, 
$W_{0,1} \sim 0.35$ including all meson resonances. The same considerations
made for $T = 0.9\ T_c$ and regarding this comparison also apply here.

\subsection{$T = T_c$}
\begin{figure}[!ht]
\includegraphics*[width=1.0\columnwidth]{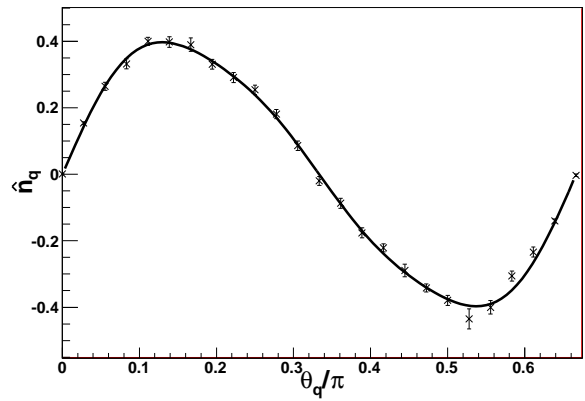}
\vspace{-0.cm}
\caption{Normalized quark density at $T = T_c$ and $\theta_I = 0$.
The solid line corresponds to the three-sine fit reported in 
Table~\ref{tabSINfit}.}
\label{figSINfitASSEb1} 
\vspace{-0.cm}
\end{figure}

Finally let us briefly discuss results obtained at $T = T_c$. Since at
this temperature we stay in the confined phase as we switch an imaginary 
chemical potential, however small, it is still sensible to test
predictions from the HRG model. However it is sufficient to look at
results obtained for $\hat{n}_q$ at $\theta_I = 0$ (Fig.~\ref{figSINfitASSEb1})
to realize that violations to the model are important: in this case inclusion
of the first three harmonics ($B = 1,2,3$) is necessary to 
obtain a reasonable value for $\tilde\chi^2$ (see Table~\ref{tabSINfit}).
This fact is confirmed by fits to the complete set of data for
$\hat{n}_q$ and $\hat{n}_I$ which are reported in 
Table~\ref{tabHRGfit}: the $\tilde\chi^2$ value decreases as
more and more terms in the expansion in Eq.~(\ref{freelowT}) are added.

In conclusion, within the current precision of our data, corrections 
to the HRG model are clearly detectable starting from $T \sim 0.951\, T_c$. 

Our best fits reported in Tab.~\ref{tabHRGfit}, which are marked 
in the $\tilde\chi^2$ field by a *, provide us with 
a valid parametrization of the free energy (apart from a constant
term).
We shall make use of these parametrizations 
in the following Sections to derive generalized susceptibilities
at $\mu_I = \mu_q = 0$ and to study the analytic continuation of 
the average phase of the fermionic determinant. Systematic
effects are expected at $T = T_c$, where the 
$\tilde\chi^2$ of our best fit is somewhat bigger than 1.

In order to check for systematic effects related to the choice of the 
interpolating function we have also performed polynomial fits 
in a limited range of chemical potentials: our results are reported in 
Table~\ref{tabPOLfit}. Fits chosen for analytic 
continuation are again
marked by a * in the $\tilde\chi^2$ field.

\section{Results at $T > T_c$ \label{formaREGIONEbuona}}

The range of imaginary chemical potentials available for 
analytic continuation is limited, above $T_c$, either by the presence 
of unphysical phase transitions related to center group dynamics
(RW transitions) or by transitions corresponding to 
the analytic continuation of the deconfinement surface
present at real chemical potentials. A full account of the high
temperature phase
structure in presence of two different imaginary chemical potentials
will be given elsewhere~\cite{delmansan}; in the present context we
are just interested in the location of such transitions for the 
two temperatures explored, i.e. $T = 1.048\, T_c$ and $T = 1.25\, T_c$.
To that aim we have performed preliminary simulations on a small
$8^3 \times 4$ lattice to get a rough idea of the phase structure 
at these temperatures and thus delimit a safe region for analytic
continuation, where to perform simulations on the larger 
$16^3 \times 4$ lattice.

\begin{figure}[!ht]
\includegraphics*[width=1.0\columnwidth]{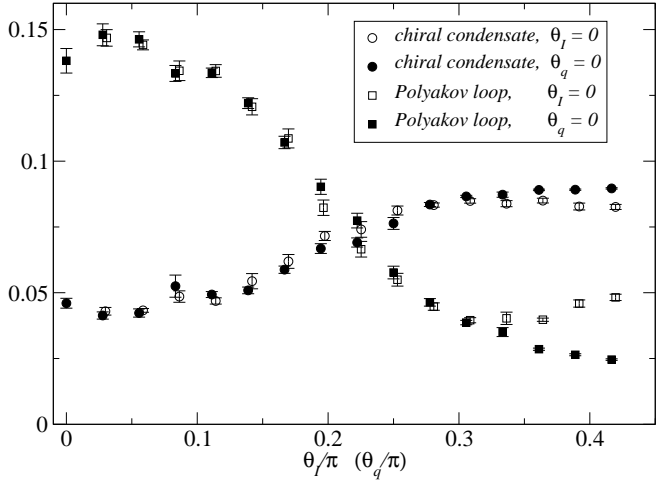}
\vspace{-0.cm}
\caption{Polyakov loop modulus and chiral condensate at $T=1.048\, T_c$ along
$\theta_I = 0$ and $\theta_q = 0$ axes. The chiral condensate has been
divided by a factor 4 to better fit in the figure.}
\label{figTRANS1.048} 
\vspace{-0.cm}
\end{figure}

As for $T = 1.048\, T_c$, in Fig.~\ref{figTRANS1.048} 
we show the behaviour of the modulus of the Polyakov loop and of the 
chiral condensate 
as a function of $\theta_q$ at $\theta_I = 0$ and as a function of
$\theta_I$ at $\theta_q = 0$. It is clear that along both axes a transition
is met where the system gets back into a phase with confinement
and chiral symmetry breaking: at those points the system 
is crossing the analytic continuation of the pseudo-critical 
surface, present also at real chemical potentials. On the same symmetry grounds
as for the deduction of general properties of the free energy 
in Section~\ref{gensusc}, one expects that
for small chemical potentials such pseudo-critical surface 
must be of the form
$$T_c (\theta_q,\theta_I) \simeq T_c (0,0) + A \theta_q^2 + B
\theta_I^2 \, .$$
As clear from Fig.~\ref{figTRANS1.048} the transition happens at approximately
equal points along both axes ($\theta_q \sim \theta_I \sim 0.2\ \pi$), 
i.e. $A \sim B$. Also the observables (Polyakov loop and chiral
condensate) seem to be, within a good approximation, universal 
functions of $| \vec \theta |$, where 
$\vec \theta \equiv (\theta_q,\theta_I)$, at least not too far from the
origin  $\theta_q = \theta_I = 0$.
Only imaginary chemical potentials strictly within the deconfined
region can be considered for analytic continuation: 
Fig.~\ref{figTRANS1.048} suggests us to take $|\vec \theta| < |\vec
\theta|_{max}$, with $|\vec\theta|_{max} \sim 0.12\ \pi$. 

\begin{figure}[!t]
\includegraphics*[width=1.0\columnwidth]{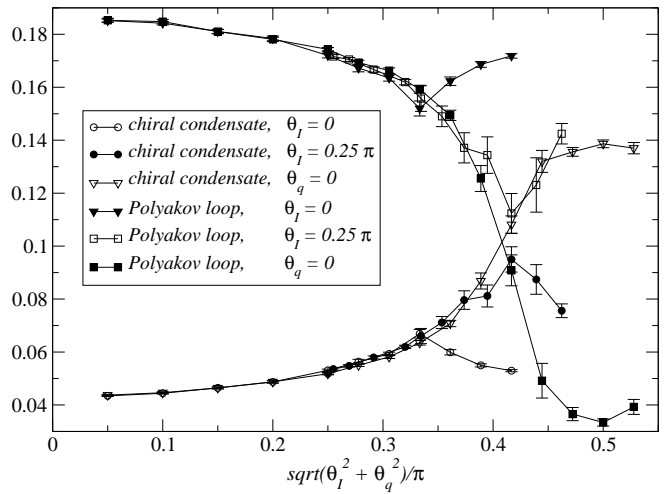}
\vspace{-0.cm}
\caption{Polyakov loop modulus and chiral condensate at $T=1.25\, T_c$
as a function of $|\vec\theta|$ and at different constant values of 
$\theta_I$ or $\theta_q$. The chiral condensate has been
divided by a factor 2 to better fit in the figure.}
\label{figobs1.25} 
\vspace{-0.cm}
\end{figure}

\begin{figure}[!t]
\includegraphics*[width=1.0\columnwidth]{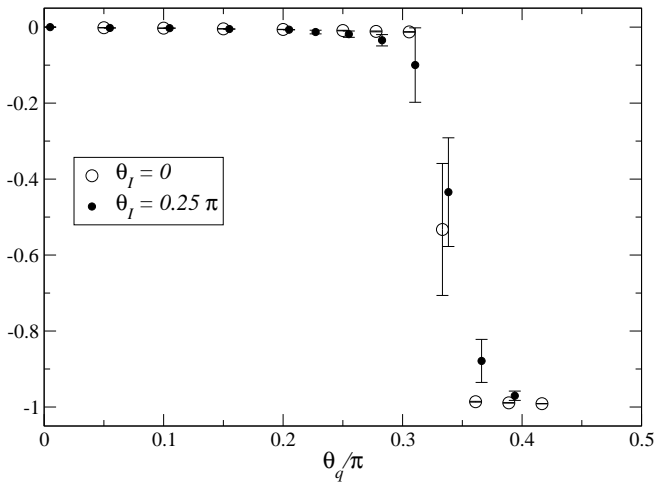}
\vspace{-0.cm}
\caption{Average phase of the Polyakov loop given in units of $2 \pi
  /3$ at $T = 1.25\, T_c$.}
\label{figphase1.25} 
\vspace{-0.cm}
\end{figure}

The phase structure is less trivial at $T = 1.25\, T_c$.
In Fig.~\ref{figobs1.25} we plot the behaviour of the modulus of
the Polyakov loop and of the chiral condensate as a function of
$|\vec\theta|$ in three cases: fixed $\theta_q = 0$, fixed $\theta_I =
0$ and fixed $\theta_I = 0.25 \pi$. We observe again an approximate universal
dependence on $|\vec\theta|$ for relatively small values of this
variable. Along the $\theta_q = 0$ axis a transition is met,
at $\theta_I \sim 0.42\ \pi$, which clearly belongs to the
pseudocritical confinement/deconfinement surface. Along the $\theta_I
= 0$ axis instead the system always stays in the deconfined phase and
the Roberge-Weiss transition is met at $\theta_q = \pi/3$
where the system enters a different $Z_3$ sector, as also apparent 
from the behaviour of the Polyakov loop phase shown in 
Fig.~\ref{figphase1.25}. What happens along the $\theta_I = 0.25 \pi$
axis is less clear: presumably there one meets a pseudo-critical
point close to the junction between the Roberge-Weiss transition and
the pseudo-critical deconfinement surface. In this context we are only 
interested in delimiting a region safe for analytic continuation: 
from Fig.~\ref{figobs1.25}
it is clear that points with $|\vec \theta| < |\vec \theta|_{max} \sim
0.3\ \pi$ are surely contained in that region
and this has been our conservative choice.

In this temperature regime we have tried to fit our results 
for $\hat{n}_q$ and $\hat{n}_I$
as a function of $\theta_q,\theta_I$ according to polynomials
derived from the general expansion for the free energy given in 
Eq.~(\ref{freehighT1}) and truncated to a given order, 
or according to expressions derived from a parametrization
of the free energy given in terms of ratios of polynomials 
as in Eq.~(\ref{freehighT2}).

At $T=1.048\, T_c$ a fourth order polynomial provides a good
fit, while coefficients are largely indetermined if a sixth order
polynomial is used: not enough information can be extracted from
the limited region available for analytic continuation. A marginally 
good fit is obtained with the ratio of two second order polynomials,
but a fourth order polynomial at the numerator seems preferable.

At $T=1.25\,T_c$ a sixth order polynomial or the ratio
between fourth and second order polynomial are instead the best
interpolating functions.

\begin{figure}[!ht]
\includegraphics*[width=1.0\columnwidth]{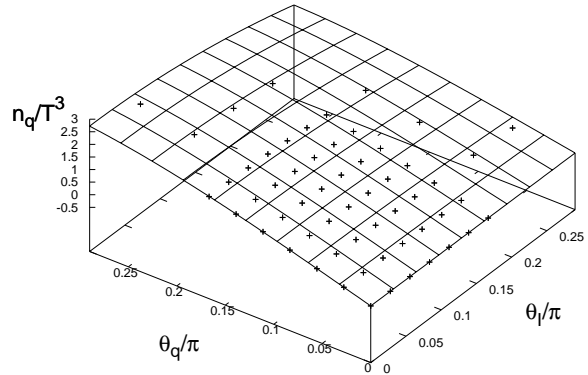}
\vspace{-0.cm}
\caption{Fit of normalized quark densities at $T = 1.25\, T_c $, obtained
from all imaginary chemical potentials in the region 
$|\vec{\theta}|\leq 0.30\pi$ (cross points),
with a sixth order polynomial function (grid surface)}
\label{figPOLfitTOT1.25} 
\vspace{-0.cm}
\end{figure}

\begin{figure}[!ht]
\includegraphics*[width=1.0\columnwidth]{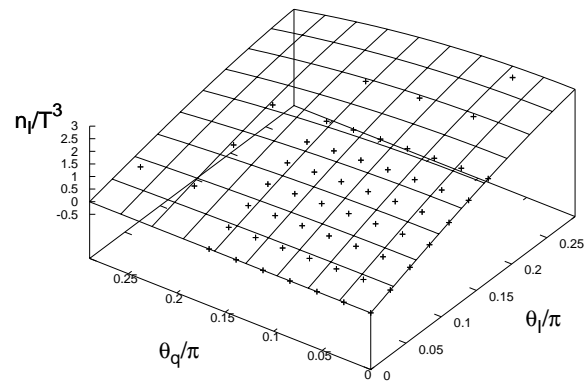}
\vspace{-0.cm}
\caption{Fit of normalized isospin densities at $T = 1.25\, T_c $, obtained
from all imaginary chemical potentials in the region 
$|\vec{\theta}|\leq 0.30\pi$ (cross points),
with a sixth order polynomial function (grid surface)}
\label{figPOLfitDIF1.25} 
\vspace{-0.cm}
\end{figure}

A complete collection of our fit results is given in Table~\ref{tabPOLfit} 
and in Table~\ref{tabRATfit}. Best fits chosen for analytic
continuation are marked again by a *.
Data obtained for $T = 1.25\, T_c$ are shown
in Figs.~\ref{figPOLfitTOT1.25} and \ref{figPOLfitDIF1.25}.\\

\section{Generalized susceptibilities}

Best fits to our data provide us with a parametrization for the
dependence of the free energy on the chemical potentials, from which 
generalized susceptibilities can be extracted. Results can be
considered reliable as long as different interpolations provide
consistent results. 

\begin{figure}[ht]
\includegraphics*[width=1.0\columnwidth]{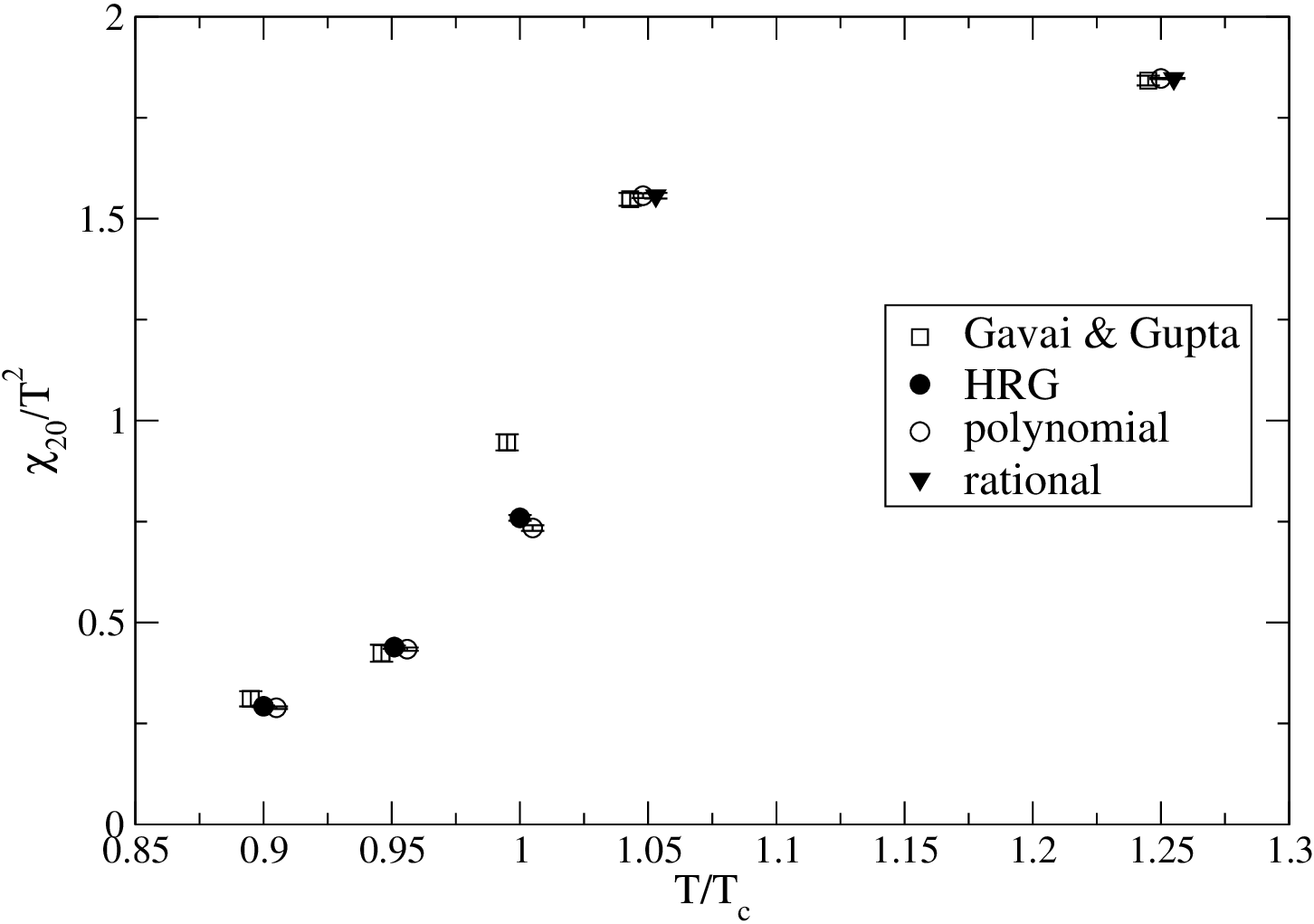}
\vspace{-0.cm}
\caption{Values obtained for $\chi_{20}/T^2$ from various fits 
and compared with results from Ref.~\cite{gagu2}.}
\label{figSUSC20} 
\vspace{-0.cm}
\end{figure}

\begin{figure}[ht]
\includegraphics*[width=1.0\columnwidth]{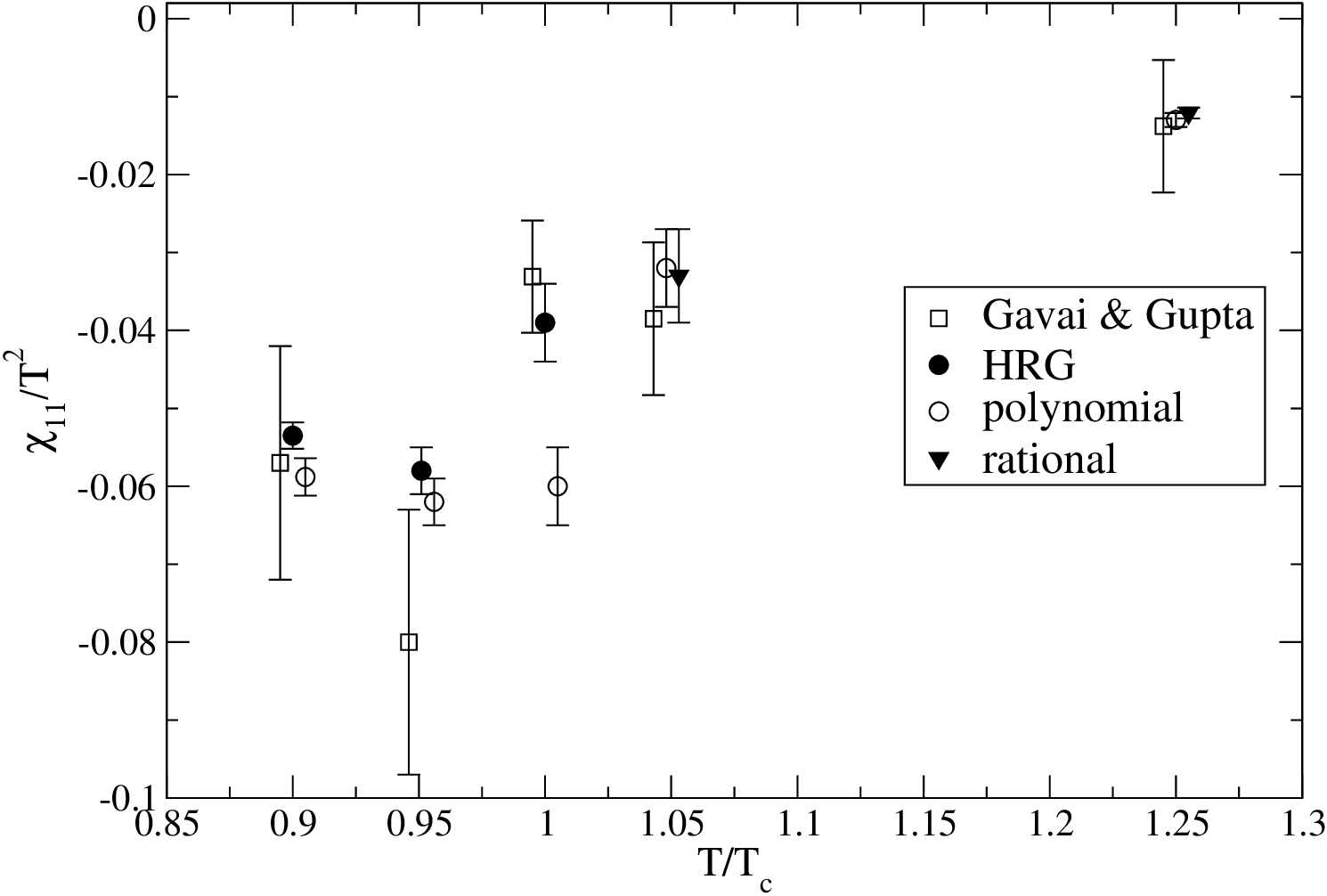}
\vspace{-0.cm}
\caption{Values obtained for $\chi_{11}/T^2$ from various fits 
and compared with results from Ref.~\cite{gagu2}.}
\label{figSUSC11} 
\vspace{-0.cm}
\end{figure}

\begin{figure}[ht]
\includegraphics*[width=1.0\columnwidth]{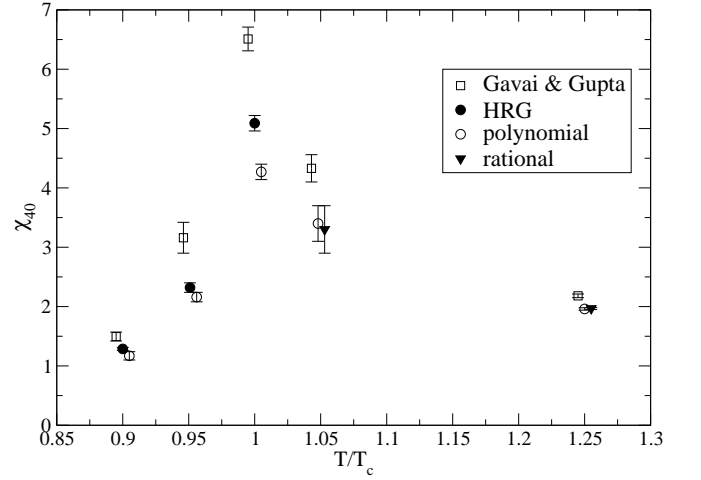}
\vspace{-0.cm}
\caption{Values obtained for $\chi_{40}$ from various fits 
and compared with results from Ref.~\cite{gagu2}.}
\label{figSUSC40} 
\vspace{-0.cm}
\end{figure}

In Table~\ref{tabSUSCall} we report results obtained for $\chi_{2,0}$,
 $\chi_{1,1}$, $\chi_{4,0}$ and $\chi_{6,0}$ (definined in 
Eq.~(\ref{gensusc2}))
from free energy best fits
marked by a * in the tables.
Results obtained for  $\chi_{2,0}$, $\chi_{1,1}$ and $\chi_{4,0}$ 
are reported also in 
Figs.~\ref{figSUSC20}, \ref{figSUSC11} and \ref{figSUSC40}
respectively,
where they are compared with analogous results obtained using the 
Taylor expansion method in Ref.~\cite{gagu2}.

The following general features can be observed. Different
extrapolations provide always consistent results for $\chi_{2,0}$
and $\chi_{1,1}$. A good agreement with Taylor expansion results
can be observed as well, apart from the $T = T_c$ case.

For $\chi_{4,0}$ we observe a discrepancy between different
interpolations only for $T = T_c$ and $T = 1.048\ T_c$; 
the agreement with Taylor expansion is less good around $T_c$.

For $\chi_{6,0}$ different extrapolations disagree or are at most 
marginally compatible in the whole range of temperatures:
with the current precision of our data, we cannot get reliable 
results for sixth or higher order susceptibilities.

In general, the comparison among different interpolation methods
and with Taylor expansion results is good, apart from the 
region around $T_c$. This in not unexpected: 
right above
$T_c$ the region of imaginary chemical potentials usable for analytic 
continuation is small and restricted by the continuation of the
pseudo-critical line, so that poor information is available.
Moreover, right at $T = T_c$ we could not get best fits
to the free energy dependence with a $\chi^2/{\rm d.o.f}$ less
than 1.5, therefore we do not have a completely satisfactory
parametrization of the free energy for this temperature and
systematic effects related to analytic continuation may be 
more important.

We have reported in Table~\ref{tabPARsim} the total number
of Dirac matrix multiplications needed in our numerical simulations
at each temperature. 
We infer, from a rough estimate, that  
the effort for measurement purposes in our case (which is 
more or less half of the total) is approximately two orders of magnitude
larger than what needed (again for measurement purposes) in 
Ref.~\cite{gagu2}. The increased effort leads to corresponding smaller 
errors (about one order of magnitude) 
only for the lowest
susceptibilities ($\chi_{20}$ and $\chi_{11}$), while for higher
order susceptibilities the Taylor expansion method seems to be more 
efficient. One has to consider, however, that our numerical 
simulations were not designed to be optimized for the computation
of susceptibilities, and that in our case we obtain a complete
parametrization of the free energy dependence in terms of 
$\mu_1$ and $\mu_2$, which is usable for different purposes.

\begin{figure}[ht]
\includegraphics*[width=1.0\columnwidth]{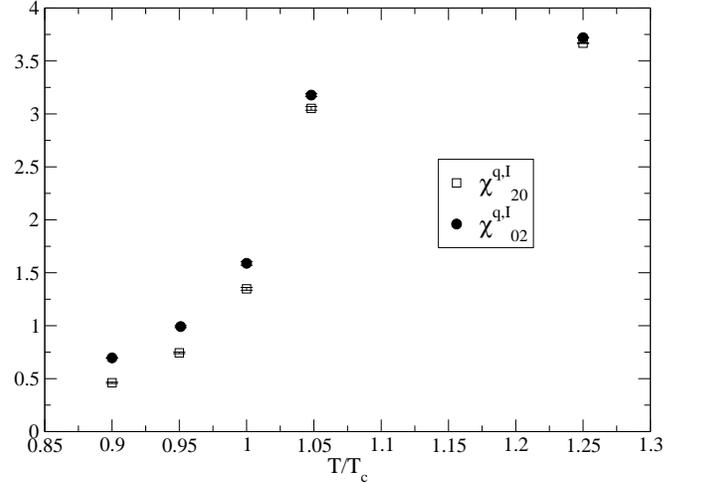}
\vspace{-0.cm}
\caption{Values obtained for $\chi_{2,0}^{q,I}/T^2$  and 
$\chi_{0,2}^{q,I}/T^2$ from polynomial fits.
}
\label{chiqI} 
\vspace{-0.cm}
\end{figure}

In Table~\ref{tabSUSCallQI} we report also results obtained 
for the susceptibilities with respect to quark and isospin chemical
potentials and defined in Eq.(\ref{gensusc2}). In Fig.~\ref{chiqI}
we show in particular the values of $\chi_{2,0}^{q,I}$  and
$\chi_{0,2}^{q,I}$ for all temperatures, as obtained from polynomial
fits: notice that $\chi_{0,2}^{q,I}$ is always larger than
$\chi_{2,0}^{q,I}$ below $T_c$, meaning that isospin charge fluctuations
can be excited more easily (mainly in the form of pions) 
than baryon charge fluctuations below $T_c$, while in the deconfined region
the two susceptibilities become almost equal, as appropriate for 
a system made up mostly of quark-like degrees of freedom.


\begin{table}
\begin{center}
\begin{tabular}{|c||c||c|c|c|c|}
\hline 
$T/T_c$ & $Fit$ & $\chi_{2,0}$ & $\chi_{1,1}$ & $\chi_{4,0}$ & $\chi_{6,0}$\\
\hline
\hline 
0.9
 & $HRG$ & 0.2925(20) & -0.0535(17)  & 1.287(24)  & 9.5(3)   \\
\hline
 & $POL$ & 0.289(3)   & -0.0588(24)  & 1.17(7)  & 5.6 $\pm$ 1.2 \\
\hline
 & \cite{gagu2} & 0.311(19)   & -0.057(15)  & 1.495(75)  & 11.2 $\pm$ 7.0 \\
\hline
\hline 
0.951
 & $HRG$ & 0.439(4) & -0.058(3) & 2.32(8)   & 22(2) \\
\hline
 & $POL$ & 0.434(4) & -0.062(3) & 2.16(8)   & 14(2) \\
\hline
 & \cite{gagu2} & 0.423(21)   & -0.080(17)  & 3.16(26)  & -29 $\pm$ 11 \\
\hline
\hline 
1
 & $HRG$ & 0.759(7) & -0.039(5) & 5.09(13) & 61(3)\\
\hline
 & $POL$ & 0.734(7) & -0.060(5) & 4.27(13) & 31(2)\\
\hline
 & \cite{gagu2} & 0.946(20)   & -0.0331(72)  & 6.51(20)  & -5.3 $\pm$ 10.7 \\
\hline
\hline 
1.048
 & $POL$ & 1.557(6) & -0.032(5) & 3.4(3) & -  \\
\hline
 & $RAT$ & 1.557(7) & -0.033(6) & 3.3(4) & 1 $\pm$ 24  \\
\hline
 & \cite{gagu2} & 1.55(16)   & -0.0385(98)  & 4.33(23)  & -69 $\pm$ 16 \\
\hline
\hline 
1.25
 & $POL$ & 1.8470(12) & -0.0130(9) & 1.960(20) & 0.64(23) \\
\hline
 & $RAT$ & 1.8473(11) & -0.0121(7) & 1.968(16) & 2.78(25) \\
\hline
 & \cite{gagu2} & 1.84(12)   & -0.0138(85)  & 2.181(31)  & 5.5 $\pm$ 1.7 \\
\hline
\end{tabular}
\caption{Table of different susceptibilities obtained from various
fits. We present the values obtained from ``best fits'' of each kind of 
free energy form, together with values obtained by the
authors of Ref.~\cite{gagu2} using the Taylor expansion method.
\label{tabSUSCall} }
\end{center}
\end{table}

\begin{table}
\begin{center}
\begin{tabular}{|c||c||c|c|c|c|c|}
\hline 
$T/T_c$ & $Fit$ & $\chi_{2,0}^{q,I}$ & $\chi_{0,2}^{q,I}$ & $\chi_{4,0}^{q,I}$ & $\chi_{0,4}^{q,I} $ & $\chi_{2,2}^{q,I} $\\
\hline
\hline 
0.9
 & $HRG$ & 0.478(6) & 0.692(4) & 4.92(25) & 4.05(7) & 1.94(3)  \\
\hline
 & $POL$ & 0.461(6) & 0.696(9) & 4.15(17) & 4.1(4)  & 1.76(14) \\
\hline
\hline
0.951
 & $HRG$ & 0.762(9) & 0.993(10) & 8.2(3)   & 8.3(5) & 3.45(11) \\
\hline
 & $POL$ & 0.744(8) & 0.992(10) & 6.95(23) & 7.5(4) & 3.36(15)\\
\hline
\hline 
1
 & $HRG$ & 1.440(14) & 1.597(16) & 19.8(5) & 16.8(8) & 7.47(21) \\
\hline
 & $POL$ & 1.348(13) & 1.589(18) & 13.9(3) & 15.6(8) & 6.46(23) \\
\hline
\hline
1.048
 & $POL$ & 3.052(17) & 3.178(15) & 7.5 $\pm1.9$ & 12.8$\pm 1.9$ & 5.5(4) \\
\hline
 & $RAT$ & 3.045(21) & 3.176(15) & 2(5) & 11(4) & 6.6(9) \\
\hline
\hline 
1.25
 & $POL$ & 3.668(3) & 3.720(3) & 4.59(13) & 4.75(12) & 3.67(3) \\
\hline
 & $RAT$ & 3.671(3) & 3.7188(24) & 4.72(11) & 4.68(9) & 3.681(17) \\
\hline
\end{tabular}
\caption{Table of different susceptibilities calculated with respect to the 
quark and isospin chemical potentials from the same best fits as for 
Tab.~\ref{tabSUSCall}
\label{tabSUSCallQI}}
\end{center}
\end{table}

\section{Phase of the fermionic determinant}
\label{phase}

As we have recalled in Section II, the 
complex phase of the fermion determinant, 
$\det M[U,\mu] = |\det M[U,\mu]| e^{i \phi}$, 
hinders numerical simulations in presence of a 
real baryon chemical potential $\mu_B$. 
The problem is however milder in case the 
fluctuations of the phase $\phi$ around zero,
over the gauge configurations which are typical 
of the statistical ensemble, are small:
in that case efficient numerical methods, like reweighting, can be used.
A typical measure of the severeness of the
sign problem is therefore given by the average of the phase factor 
(or some power of it), computed for convenience over the ensemble 
at finite isospin density. In particular in our
case we can define:
\beq
\langle e^{i \phi/2} \rangle_\mu  &\equiv& 
\left\langle {\det M^{1 \over 4} (\mu) \over \det M^{1\over 4} (-\mu)}
\right\rangle_{(\mu,-\mu)} = {Z(\mu,\mu) \over Z(\mu,-\mu)} \nonumber \\
&=& {Z(\mu_q = \mu,\mu_I = 0) \over Z(\mu_q = 0,\mu_I = \mu)} 
\, .
\label{phase1}
\eeq

As clear from Eq.~(\ref{phase1}), a way to determine 
$\langle e^{i \phi/2} \rangle_\mu$ is to take 
the average of the ratio of two determinants over
the ensemble at real isospin chemical potential: that is 
feasible but computationally demanding, especially at large
volumes. Studying the analytic continuation of 
$\langle e^{i \phi/2} \rangle_\mu$ at imaginary values
of $\mu$,
\beq
\langle e^{i \phi/2} \rangle_{i \mu}  &\equiv& 
{Z(i \mu,i \mu) \over Z(i \mu,- i \mu)}\, , 
\label{phase2}
\eeq
is an alternative: 
it has been shown~\cite{splitt1,splitt2} that, in the full QCD 
case, the average phase factor is analytic around $\mu^2 =0$, 
and an efficient numerical method for the evaluation of the ratio
of partition functions appearing in Eq.~(\ref{phase2}) has been
proposed in Ref.~\cite{conradi}. 

In the present context we 
adopt a much faster and cheaper approach: having measured and fitted 
first derivatives with respect to both chemical potentials, we have a 
complete knowledge, apart from constant terms, of the dependence of the 
free energy on $\mu_1$ and $\mu_2$,
so that computing the ratio in Eq.~(\ref{phase2}) is straightforward.
Let us consider for instance the low temperature case, where we have
used the HRG parametrization in Eq.~(\ref{imF}) that we rewrite:
\beq
F = - V T^4 &&\sum_{B,I}  W_{B,I}(T) 
\bar\delta(B)
\cos ( 3 B \theta_q) \nonumber \\ && \left( \sum_{I_3 \geq 0} 
\bar\delta(I_3)\cos ( 2 I_3 \theta_I ) \right)
\label{imF2}
\eeq
then 
\beq
\frac{Z(\theta_q = \theta,\theta_I = 0)}{Z(\theta_q = 0,\theta_I =
  \theta)} = e^{-\frac{1}{T} \left(
F(\theta_q = \theta,\theta_I = 0) - F(\theta_q = 0,\theta_I = \theta)
  \right)} 
\eeq
Non-zero coefficients $W_{B,I}$, apart from the constant
$W_{0,0}$ which does not enter in the computation of the 
average phase factor, have been obtained by fitting our numerical
data. The expression can then be easily continued to 
real chemical potentials obtaining:
\beq
\langle e^{i \phi/2} \rangle_{\mu} &=& 
 \exp \left( \frac{N_s^3}{N_t^3} 
\sum_{B,I}  W_{B,I}(T) 
\bar\delta(B)
\left( \cos ( 3 B \theta_q)  - \right.\right.
\nonumber \\
&& \left. \sum_{I_3 \geq 0} 
\bar\delta(I_3)\cos ( 2 I_3 \theta_I ) )
\right) \, .
\label{HRGphase}
\eeq

The same procedure applies to other functional forms used in our fits: the
comparison of different extrapolations to real chemical potentials
based on different fitting functions, when available, gives a measure of the
systematic effects involved in analytic continuation. Notice that in
the case of the HRG parametrization we can distinguish the
different contributions to the average phase factor, hence to the sign
problem, coming from different particle species: this feature will be useful 
in our analysis.

\begin{figure}[ht]
\includegraphics*[width=1.0\columnwidth]{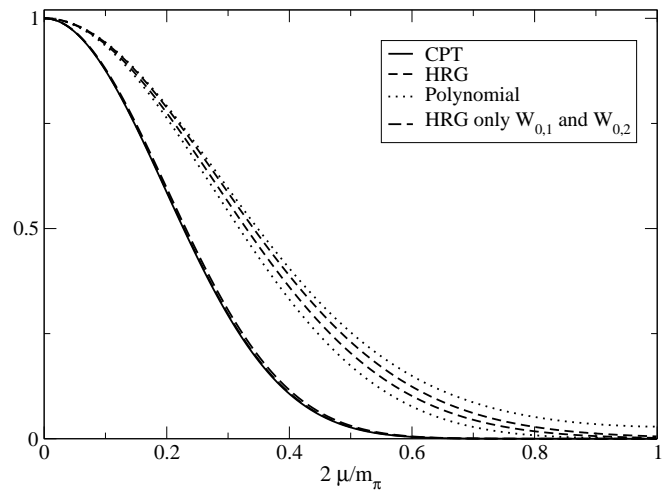}
\vspace{-0.cm}
\caption{The average phase factor continued from different
  interpolations and compared to 1-loop $\chi$PT results for $T = 0.9\
  T_c$. In particular we show the 90\% confidence level band
  extrapolated from our best fits to the free energy dependence.}
\label{figfase0.9} 
\vspace{-0.cm}
\end{figure}

\begin{figure}[t]
\includegraphics*[width=1.0\columnwidth]{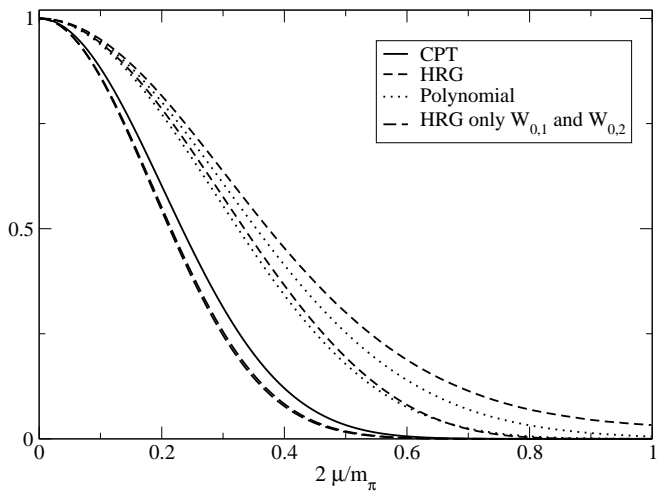}
\vspace{-0.cm}
\caption{Same as in Fig.~\ref{figfase0.9} for $T = 0.951\ T_c$.}
\label{figfase0.95} 
\vspace{-0.cm}
\end{figure}

In Figs.~\ref{figfase0.9} and \ref{figfase0.95} we report,  
as a function of $2 \mu/ m_\pi$, results 
obtained respectively at $T = 0.9\ T_c$ and $T = 0.951\ T_c$
using HRG inspired and polynomial interpolations. Where visible, the
two lines reported for each extrapolation delimit the 90\% confidence
level region and give an estimate of our 
uncertainties: a good agreement between HRG inspired and polynomial 
extrapolations can be appreciated.

It is interesting to make a direct comparison of our results with
predictions coming from chiral perturbation theory ($\chi$PT). The average phase
factor has been computed to one loop order of $\chi$PT in
Ref.~\cite{splitt3}.
According to the results reported in Section VI of
Ref.~\cite{splitt3}, our spatial lattice size is big enough ($L_s m_\pi \sim
6.6$) to justify taking the thermodynamical limit at fixed T of the 
one loop $\chi$PT result, which coincides with the prediction of a HRG
model including only pions: 
\beq
 \langle e^{i \phi/2} \rangle_{\mu}  = e^{-\Delta G_0}
\label{CPTHRG1}
\eeq
with
\beq
\Delta G_0 = V T^3
\left( \frac{m_{\pi}}{T\pi} \right)^2\sum_{n=1}^{+\infty}\frac{K_2
\left( \frac{n m_{\pi}}{T} \right)}{n^2}
\left( \cosh(2 \mu n) -1 \right) 
\label{CPTHRG2}
\eeq
This prediction (assuming in our case $m_\pi \simeq 280$ MeV and 
$T_c \simeq 170$ MeV) is reported in Figs.~\ref{figfase0.9} and 
\ref{figfase0.95} as a solid line. It is apparent that the 
agreement of $\chi$PT with the analytic continuation of our
data is not satisfactory. In particular analytic continuation provides
a higher value for  $\langle e^{i \phi/2} \rangle_{\mu}$, meaning 
a milder sign problem. To better understand
the origin of this discrepancy, we have tried to compute the average 
phase factor from our HRG model best fit, but neglecting all 
contributions to the free energy with $B \neq 0$, which cannot
be taken into account by $\chi$PT, i.e. taking only contributions from
$W_{0,1}$ and $W_{0,2}$ in Eq.~(\ref{HRGphase}). 
Results are shown in Figs.~\ref{figfase0.9} and 
\ref{figfase0.95}: in this case the agreement with $\chi$PT is almost perfect 
for $T = 0.9\ T_c$, and acceptable for $T = 0.951\ T_c$.
This is expected since, as we have discussed in Section~\ref{T0.9},
the coefficients $W_{0,1}$ and $W_{0,2}$ obtained by our fits are 
compatible within errors, at $T = 0.9\ T_c$,
with those predicted if only pions are taken into account: of course
that may be an accident and the contribution of higher meson
resonances should be better understood.

Anyway, an outcome of our analysis, which is in agreement with HRG
model expectations, is that contributions to the average phase factor coming 
from physical states with $B \neq 0$ are significant and tend in
general to make the sign problem less severe. 

\begin{figure}[t]
\includegraphics*[width=1.0\columnwidth]{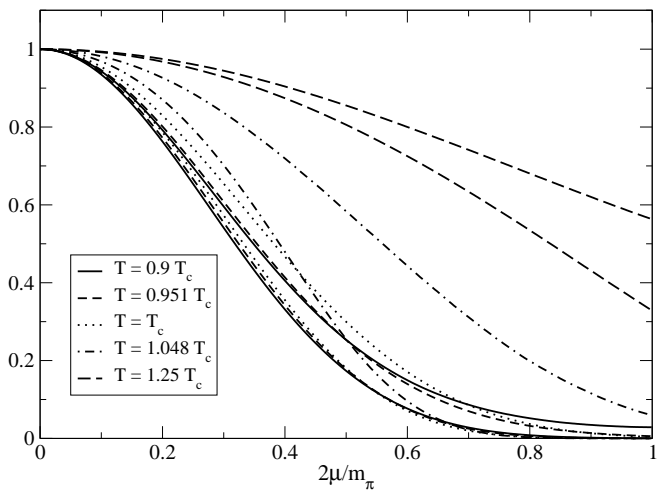}
\vspace{-0.cm}
\caption{The average phase factor continued from polynomial
  interpolations at all explored temperatures. For each temperature
we show the 90\% confidence level band corresponding to our best fits.}
\label{figfasetot} 
\vspace{-0.cm}
\end{figure}

In Fig.~\ref{figfasetot} we report the analytic continuation of the
average phase factor obtained at all temperatures from a polynomial
fit: of course results reported in the figure must be intended to be valid 
for chemical potentials bounded, below $T_c$, 
by the deconfinement critical line present at real chemical potentials.
As expected, at fixed chemical potential the sign problem is much milder
for $T > T_c$. This can be put again in connection with the fact
that states with $B \neq 0$, which are more easily created above
$T_c$, tend to mitigate the sign problem.

\section{Conclusions}
\label{conclusions}

In this paper we have studied $N_f = 2$ QCD thermodynamics, exploiting
analytic continuation from two imaginary chemical
potentials coupled to baryon and isospin charges. Simulations
have been performed at five temperatures around the critical 
value $T_c \simeq 170$ MeV, using 
a $16^3 \times 4$ lattice with a standard staggered action 
and a fixed pion mass $m_\pi \simeq 280$ MeV.

We have computed free energy first derivatives with respect to 
the chemical potentials (quark number densities) and interpolated them
by suitable functions, in order to perform analytic continuation. 
In particular we have tested HRG predictions below $T_c$,
reconstructed generalized susceptibilities at zero chemical potentials
and determined the analytic continuation of the average phase factor.

We have checked that HRG model predictions are in very good agreement
with our numerical results for $T = 0.9\ T_c$. Small but clearly detectable
deviations start to be visible at $T = 0.95\ T_c$, in agreement with
similar results reported in Ref.~\cite{taylor2}.
They appear, in a HRG inspired parametrization of the free energy,  
as contributions from unphysical states with 
higher values of baryon or isospin charges, which are of the order of
a few percent at $T = 0.95\ T_c$ and above 10 \% at $T \simeq T_c$.

Regarding the computation of generalized susceptibilities, analytic 
continuation gives consistent results which are in
agreement with those obtained by the Taylor expansion method, apart 
from temperatures in correspondence or right above $T_c$,
where the range of imaginary chemical potentials available for
analytic continuation is small and larger systematic effects are expected. 
Poor information has been obtained for susceptibilities beyond sixth order.

We have obtained consistent determinations, by analytic continuation
with different interpolating functions, of the average phase factor.
In particular below $T_c$, 
in the case of HRG inspired interpolations, we have been able to distinguish
the contribution to the average phase factor coming from the different
hadron states: results from analytic continuation are consistent 
with $\chi$PT results, below $T_c$, if one takes into account only meson 
contributions. Baryons
give contributions to the average phase factor which in general
tend to make the sign problem less severe.
The sign problem is much milder
for $T > T_c$, and this can be put again in connection with the fact
that states with $B \neq 0$, which are more easily created above
$T_c$, tend to mitigate the sign problem.

Our results should be refined and could be improved in several
respects. Simulation closer to the continuum limit and possibly
closer to the physical quark mass spectrum would clarify the comparison
with HRG predicitions, as well as that with $\chi$PT for the average phase
factor. An improvement in the determination of generalized susceptibilities
could be obtained by combining analytic continuation with other
techniques: for instance fixing lowest order terms in a polynomial
expansion by the Taylor expansion method or by reweighting could 
lead to enhanced predictivity for analytic continuation. We shall
continue our investigation along those lines in the future.

\section*{Acknowledgments}
We thank F.~Becattini, F.~Karsch and K.~Splittorf 
for very useful discussions, as well as 
R.~Gavai and S.~Gupta for very interesting discussions and for providing
us with their numerical results for generalized non-linear susceptibilities.
Numerical simulations have been performed on two PC farms 
in Genoa and in Bari provided by INFN.


\begin{table*}
\begin{center}
\begin{tabular}{|c|c|c||c|}
\hline 
$c_1$ & $c_2$ & $c_3$ & $\chi^2/d.o.f$\\
\hline
\hline 
\multicolumn{4}{|c|}{$T=0.9~T_c$}\\ 
\hline
 0.1536(14) & -          & -    &
  28/21 \\
\hline 
 0.1514(15) & 0.0046(14) & -    &
 17/20 \\
\hline 
\hline
\multicolumn{4}{|c|}{$T=0.951~T_c$}\\ 
\hline
 0.2413(25) & -          & -    &
 42/21 \\
\hline 
 0.2383(26) & 0.0102(22) & -    &
 21/20 \\
\hline
\hline 
\multicolumn{4}{|c|}{$T=T_c$}\\ 
\hline
 0.3865(4) & -        & -        &
  248/21 \\
\hline 
 0.395(4) & 0.048(3) & -        &
  50/20 \\
\hline 
 0.389(4) & 0.048(3) & 0.018(4) &
  23/19 \\
\hline
\end{tabular} 
\caption{Coefficients of sinusoidal fits for $\hat{n}_q$ along
  $\theta_i=0$ axis at various temperatures (see Eq.~(\ref{sinfit})).
Blank columns stand for terms not included in the fits. \label{tabSINfit}}
\end{center}
\end{table*}


\begin{table*}
\begin{center}
\begin{tabular}{|c||c|c|c|}
\hline 
$T/T_c$ & $c_1$ & $c_2$ & $c_3$      \\
\hline
\hline
 0.9 &
  0.1521(11) & 0.0052(11) & -        \\
\hline 
 0.951 &
  0.2387(19) & 0.0101(17) & -        \\
\hline 
 1 &
  0.392(3)   & 0.0503(27) & 0.018(3)  \\
\hline 
\end{tabular}
\caption{Weight of differents harmonics  at various temperatures for 
 $\hat{n}_q$ at  $\theta_i=0$ (see Eq.~(\ref{sinfit})) obtained by Fourier
 transform. Blank columns correspond to terms not included in the 
 previous fits. 
\label{tabSINfourier}}
\end{center}
\end{table*}


\begin{table*}
\begin{centering}
\begin{tabular}{|c|c||c|c|c|c||c|c||c|}
\hline 
$W_{0,1}$  & $W_{0,2}$  & $W_{1,\frac{1}{2}}$  & $W_{1,\frac{3}{2}}$  & $W_{1,\frac{5}{2}}$ & $W_{1,\frac{7}{2}}$ & $W_{2,1}$  & $W_{2,2}$ & ${\chi}^{2}/{\rm d.o.f.}$\\
\hline
\hline 
\multicolumn{9}{|c|}{$T=0.9~T_c$}\\ 
\hline
 0.2284(11)  & -  & 0.0110(6)  & 0.0202(3)  & - & - & -  & - &
  284/187 \\
\hline
 0.2157(18) & 0.0050(6) & 0.0115(6) & 0.0198(3)  & - & - & -  & - &
  206/186 \\
\hline 
 0.2156(18) & 0.0051(6) & 0.0111(6) & 0.0197(3) & - & - & 0.00043(13) & - &
  196/185 *\\
\hline
\hline 
\multicolumn{9}{|c|}{$T=0.951~T_c$}\\ 
\hline
 0.2862(13) & -  & 0.0199(7) & 0.0305(4) & - & - & -  & - &
  640/187 \\
\hline
 0.258(2) & 0.0114(6) & 0.0212(7) & 0.0292(4) & - & - & -  & - &
  281/186 \\
\hline
 0.257(2) & 0.0117(6) & 0.0203(8) & 0.0290(4) & - & - & 0.00084(18) & - &
  259/185 \\
\hline
 0.256(2) & 0.0114(6) & 0.0210(8) & 0.0264(6) & 0.0017(3) & - & 0.00088(18) & - &
  230/184 \\
\hline 
 0.257(2) & 0.0106(7) & 0.0212(8) & 0.0265(6) & 0.0009(4) & 0.0006(2) & 0.00090(18) & - &
  222/183 *\\
\hline
\hline 
\multicolumn{9}{|c|}{$T=T_c$}\\ 
\hline
0.3775(15) & -          & 0.0412(7) & 0.0456(4) & - & - & -  & - &
  1798/187 \\
\hline 
 0.3322(21) & 0.0219(7) & 0.0391(7) & 0.0465(4) & - & - & -  & - &
  808/186 \\
\hline 
 0.3269(21) & 0.0225(7) & 0.0363(7) & 0.0464(4) & - & - & 0.00372(24) & - &
  562/185 \\
\hline 
 0.3184(22) & 0.0246(7) & 0.0349(7) & 0.0396(6) & 0.0053(3) & - & 0.00436(24) & - &
  330/184 \\
\hline 
 0.3208(22) & 0.0218(8) & 0.0342(7) & 0.0391(6) & 0.0038(4) & 0.0019(3) & 0.00445(24) & - &
  288/183 \\
\hline 
 0.3214(22) & 0.0220(8) & 0.0344(8) & 0.0393(6) & 0.0042(4) & 0.0015(3) & 0.0031(6) & 0.010(4) &
  281/182 *\\\hline
\end{tabular}
\par\end{centering}

\caption{Coefficients of HRG model fits at various temperatures. 
\label{tabHRGfit}}
\end{table*}


\begin{table*}
\begin{center}
\begin{tabular}{|c||c|c||c|c|c||c|c|c|c||c|}
\hline 
$|\vec{\theta}|_{max}/\pi$ & $c_{20}$ & $c_{11}$ & $c_{40}$ & 
$c_{22}$ & $c_{04}$ & $c_{60}$ & $c_{42}$ & $c_{24}$ & $c_{06}$ & 
$\chi^{2}/d.o.f$\\
\hline
\hline 
\multicolumn{11}{|c|}{$T=0.9~T_c$}\\ 
\hline
 0.34 &
  0.479(3) & 0.1892(22) & - & - & - & - & - & - & - &
 6014/108 \\
\hline
 0.34 &
  0.659(5) & 0.412(4)   & -2.66(7) & -1.22(4) & -2.40(4) & - & - & - & - &
 237/105 \\
\hline
 0.34 &
  0.696(9) & 0.461(6) & -4.1(4) & -1.76(14) & -4.15(17) & 23(6) & 8(3) & 12(3) & 30(3) &
 113/101 *\\
\hline
\hline 
\multicolumn{11}{|c|}{$T=0.951~T_c$}\\ 
\hline 
 0.34 &
  0.636(3) & 0.301(3) & - & - & - & - & - & - & - &
 8483/108 \\
\hline 
 0.34 &
  0.897(5) & 0.651(5) & -3.69(6) & -2.00(5) & -3.76(5) & - & - & - & - &
 388/105 \\
\hline 
 0.34 &
  0.992(10) & 0.744(8) & -7.5(4) & -3.36(15) & -6.95(23) & 62(7) & 28(4) & 26(4) & 54(4) &
 125/101 *\\
\hline
\hline 
\multicolumn{11}{|c|}{$T=T_c$}\\ 
\hline
 0.34 &
  0.838(5) & 0.394(4) & - & - & - & - & - & - & - & 
 12955/108 \\
\hline
 0.34 &
  1.340(9) & 1.099(9) & -5.75(8) & -3.33(8) & -6.67(7) & - & - & - & - & 
 972/105 \\
\hline
 0.34 &
  1.589(18) & 1.348(13) & -15.6(8) & -6.46(23) & -13.9(3) & 157(13) & 65(6) & 47(5) & 121(5) & 
 239/101 *\\
\hline
\hline
\multicolumn{11}{|c|}{$T=1.048~T_c$}\\ 
\hline
 0.12 & 
  3.029(8) & -2.941(9) & - & - & - & - & - & - & - & 
 228/36 \\
\hline
 0.12 & 
  3.178(15) & 3.052(16) & -12.8 $\pm$ 1.9 & -5.5(4) & -7.5 $\pm$ 1.9 & - & - & - & - & 
 39/33 *\\
\hline
 0.12 &
  3.178(24) & 3.05(3) & -10(6) & -7.6 $\pm$ 1.5 & -4(6) & 639(1005) & 125(148) & 224(135) & -991(1037) & 
 34/29 \\
\hline
\hline 
\multicolumn{11}{|c|}{$T=1.25~T_c$}\\ 
\hline
 0.3 & 
  3.2810(11) & 3.2438(12) & - & - & - & - & -& -& - & 
  170616/111 \\
\hline
 0.3 & 
  3.7156(18) & 3.6677(2) & -4.67(4) & -3.555(9) & -4.74(4) & - & -& -& - & 
 142/111 \\
\hline
 0.3 &  
  3.720(3) & 3.668(3) & -4.75(12) & -3.67(3) & -4.59(13) & 0(3) & 1.7(5) & 1.5(5) & 6(3) &
 123/107 *\\
\hline
\end{tabular}
\end{center}
\caption{Coefficients of polynomial fits at various temperatures. 
Each line contains results of fit performed on all points in the
circumference of radius $|\vec{\theta}|_{max}$
\label{tabPOLfit} }
\end{table*}


\begin{table*}
\begin{center}
\begin{tabular}{|c||c|c|c|c|c||c|c|c|c|c||c|}
\hline 
$|\vec{\theta}|_{max}/\pi$  & $n_{20}$ & $n_{02}$ & $n_{40}$ & $n_{22}$ & $n_{04}$ & $d_{20}$ & $d_{02}$ & $d_{40}$ & $d_{22}$ & $d_{04}$ & $\chi^2/d.o.f$\\
\hline
\hline 
\multicolumn{12}{|c|}{$T=1.048~T_c$}\\ 
\hline 
 0.12  &
  3.176(15) & 3.062(16) & - & - & - & 0.81(8) & 0.60(8) & - & - & - &
 55/34 \\
\hline 
 &
  3.185(23) & 3.07(3) & - & - & - & 0.9(3) & 0.7(3) & -14(17) & 3(4) & -18(18) &
 45/31 \\
\hline 
 &
  3.178(15) & 3.048(22) & 3(27) & -6(6) & -19(24) & -0.8$\pm$1.2 & -0.7$\pm$1.6 & - & - & - &
 39/31 *\\
\hline 
 &
  3.22(3) & 3.07(3) & -60(10) & -24(4) & -41(10) & -1.7(4) & -1.7(4) & -127(85) & -83(35) & -34(53) &
 32/28 \\
\hline 
\hline 
\multicolumn{12}{|c|}{$T=1.25~T_c$}\\ 
\hline 
 0.3 &
  3.7123(17) & 3.6928(19) & - & - & - & 0.3236(12) & 0.3217(11) & - & - & - &
 25705/114 \\
\hline 
 &
  3.723(3) & 3.694(3) & - & - & - & 0.329(4) & 0.353(4) & -1.58(4) & 0.589(11) & -1.66(4) &
 3941/111 \\
\hline 
 &
  3.7188(24) & 3.671(3) & -0.5(5) & -2.29(15) & -0.5(6) & 0.185(20) & 0.192(22) & - & - & - &
 124/111 *\\
\hline 
 &
  3.721(3) & 3.666(3) & -7.2(3) & -4.03(7) & -4.57(11) & -0.101(11) & -0.004(6) & -0.45(21) & -0.139(20) & 0.20(7) &
 138/108 \\
\hline 
\end{tabular}
\end{center}
\caption{Coefficients of rational fits at various temperatures. \label{tabRATfit} }
\end{table*}


\begin{thebibliography}{99}
\bibitem{glasgow}
I.~M.~Barbour, S.~E.~Morrison, E.~G.~Klepfish, J.~B.~Kogut and M.~P.~Lombardo,
 Nucl.\ Phys.\ Proc.\ Suppl.\  {\bf 60A}, 220 (1998).
\bibitem{fodor} Z.~Fodor, S.~D.~Katz, Phys.\ Lett.\ B {\bf 534} (2002) 87;
JHEP {\bf 0203}, 014 (2002).
\bibitem{density} 
Z.~Fodor, S.~D.~Katz and C.~Schmidt,
  JHEP {\bf 0703}, 121 (2007).
\bibitem{muim} Ph.~de Forcrand and O.~Philipsen, 
Nucl.\ Phys.\ B {\bf 642}, 290 (2002); Nucl.\ Phys.\  B {\bf 673}, 170 (2003).
\bibitem{immu_dl} M.~D'Elia and M.P.~Lombardo, Phys. Rev. D {\bf 67},
  014505 (2003); Phys. Rev. D {\bf 70}, 074509 (2004).
\bibitem{azcoiti}
  V.~Azcoiti, G.~Di Carlo, A.~Galante and V.~Laliena,
  Nucl.\ Phys.\ B {\bf 723}, 77 (2005).
\bibitem{chen}
  H.~S.~Chen and X.~Q.~Luo,
  Phys.\ Rev.\ D {\bf 72}, 034504 (2005).
\bibitem{giudice}
  P.~Giudice and A.~Papa,
  Phys.\ Rev.\  D {\bf 69}, 094509 (2004)
\bibitem{cea}
P.~Cea, L.~Cosmai, M.~D'Elia and A.~Papa,
  JHEP {\bf 0702}, 066 (2007).
\bibitem{sqgp}
M.~D'Elia, F.~Di Renzo and  M.P.~Lombardo, 
  Phys.\ Rev.\  D {\bf 76}, 114509 (2007)
\bibitem{conradi}
S.~Conradi and M.~D'Elia
Phys.\ Rev.\  D {\bf 76}, 074501 (2007)
\bibitem{cea2}
P.~Cea, L.~Cosmai, M.~D'Elia and A.~Papa,
Phys.\ Rev.\  D {\bf 77}, 051501 (2008)
\bibitem{rw}
  A.~Roberge and N.~Weiss,
  Nucl.\ Phys.\  B {\bf 275}, 734 (1986).
\bibitem{cano1}
S.~Kratochvila and P.~de Forcrand,
  PoS {\bf LAT2005}, 167 (2006).
\bibitem{cano2}
  A.~Alexandru, M.~Faber, I.~Horvath and K.~F.~Liu,
  Phys.\ Rev.\  D {\bf 72}, 114513 (2005).
\bibitem{taylor1} C.~R.~Allton et al., Phys. Rev. D {\bf 66}, 074507
  (2002); Phys.\ Rev.\  D {\bf 71}, 054508 (2005).
\bibitem{taylor2} 
M.~Cheng {\it et al.},
  arXiv:0811.1006 [hep-lat].
\bibitem{gagu1} 
R.~V.~Gavai and S.~Gupta,
Phys.\ Rev.\ D {\bf 68}, 034506 (2003); 
\bibitem{gagu2} 
R.~V.~Gavai and S.~Gupta,
Phys.\ Rev.\  D {\bf 71}, 114014 (2005).
\bibitem{gagu3} 
R.~V.~Gavai and S.~Gupta,
Phys.\ Rev.\  D {\bf 78}, 114503 (2008)
\bibitem{hmass1} T.~C.~Blum, J.~E.~Hetrick and D.~Toussaint,
  Phys.\ Rev.\ Lett.\  {\bf 76}, 1019 (1996).
\bibitem{hmass2}
J.~Engels, O.~Kaczmarek, F.~Karsch and E.~Laermann,
  Nucl.\ Phys.\  B {\bf 558}, 307 (1999).
\bibitem{hmass3} 
R.~De Pietri, A.~Feo, E.~Seiler and I.~O.~Stamatescu,
Phys.\ Rev.\  D {\bf 76}, 114501 (2007).
\bibitem{mpl05}
M.~P. Lombardo, {PoS} {\bf LAT2005} (2006) 168 [arXiv:hep-lat/0509181].
\bibitem{shinno}
  Y.~Shinno and H.~Yoneyama,
  arXiv:0903.0922 [hep-lat].
\bibitem{redlich}
J.~Cleymans and K.~Redlich,
  Phys.\ Rev.\  C {\bf 60}, 054908 (1999).
\bibitem{redlich2}
F.~Becattini, J.~Cleymans, A.~Keranen, E.~Suhonen and K.~Redlich,
  Phys.\ Rev.\  C {\bf 64}, 024901 (2001).
\bibitem{andronic}
A.~Andronic, P.~Braun-Munzinger and J.~Stachel,
  Nucl.\ Phys.\  A {\bf 772}, 167 (2006).
\bibitem{kareta}
F.~Karsch, K.~Redlich and A.~Tawfik,
  Phys.\ Lett.\  B {\bf 571}, 67 (2003).
\bibitem{jaffe}
R.~L.~Jaffe,
  Phys.\ Rept.\  {\bf 409}, 1 (2005).
\bibitem{splitt0}
  K.~Splittorff,
  PoS {\bf LAT2006}, 023 (2006)
  [arXiv:hep-lat/0610072].
\bibitem{splitt1}
  K.~Splittorff and J.~J.~M.~Verbaarschot,
  Phys.\ Rev.\  D {\bf 75}, 116003 (2007).
\bibitem{splitt2}
  K.~Splittorff and B.~Svetitsky,
  Phys.\ Rev.\  D {\bf 75}, 114504 (2007).
\bibitem{splitt3}
  K.~Splittorff and J.~J.~M.~Verbaarschot,
  Phys.\ Rev.\  D {\bf 77}, 014514 (2008)
  [arXiv:0709.2218 [hep-lat]].
\bibitem{PDG}
C. Amsler et al., Phys. Lett. {\bf {B667}}, 1 (2008). 
\bibitem{delmansan}
M.~D'Elia, C.~Manneschi and F.~Sanfilippo, in progress.

\end{thebibliography}
\end{document}